\documentclass[11pt]{article}

\usepackage{amsmath}
\usepackage{amsfonts}
\usepackage{amssymb}
\usepackage{multicol}
\usepackage{color}
\usepackage{graphicx}
\usepackage{pstcol}
\usepackage{pst-node}
\usepackage{pst-coil}
\usepackage{pst-plot}
\usepackage{colortbl}
\usepackage[all]{xy}
\usepackage{rotating}
\usepackage{bbm}

\newpsobject{showgrid}{psgrid}{subgriddiv=1,griddots=10,gridlabels=6pt}

\newcommand{\Show}[2]{\psshadowbox[fillstyle=solid,fillcolor=#1]{\txt{#2}}}

\newcommand{\ShowX}[3]{{\Show{#1}{\begin{minipage}[t]{#2}\center{#3}\end{minipage}}}}

\newcommand{\ShowC}[2]{\pscirclebox[shadow=true,fillstyle=solid,fillcolor=#1]{\txt{#2}}}

\newcommand{\ShowCX}[3]{{\ShowC{#1}{\begin{minipage}[t]{#2}\center{#3}\end{minipage}}}}

\newcommand{\ShowXl}[3]{{\Show{#1}{\begin{minipage}[t]{#2}{#3}\end{minipage}}}}

\definecolor{dark-green}{rgb}{0,0.7,0}
\definecolor{dark-blue}{rgb}{0,0.2,0.5}
\definecolor{med-blue}{rgb}{0,0.7,1}
\definecolor{mblue}{rgb}{0,0.2,1}
\definecolor{cnc}{rgb}{0.8,0,0}
\definecolor{light-red}{rgb}{1,0.8,0.8}
\definecolor{dark-yellow}{rgb}{1,0.8,0}
\definecolor{light-blue}{rgb}{0.8,0.9,1}
\definecolor{verylight-blue}{rgb}{0.93,0.95,1}
\definecolor{light-yellow}{rgb}{1,0.9,0.8}
\definecolor{grey}{gray}{0.88}

\begin{document}

\title{The Search for Quantum Gravity Signals}

\author{G. Amelino--Camelia$^1$, C.\ L{\"a}mmerzahl$^{2}$, A. Macias$^3$ and H. M\"uller$^4$\\
$^1$ Dipt~Fisica, Univ~La Sapienza and Sez Roma1 INFN, Ple Moro 2, Rome, Italy\\
$^2$ ZARM, University of Bremen, Am Fallturm, 28359 Bremen, Germany \\
$^3$ Departamento de F\'{\i}sica, Universidad Aut\'onoma Metropolitana-Iztapalapa, \\ 
Apartado Postal 55-534, C.P. 09340 M\'exico, D.F., Mexico \\
$^4$ Physics Department, Stanford University, Stanford, CA 94305-4060, USA}

\maketitle

\begin{abstract}
We give an overview of ongoing searches for effects motivated by the study of the quantum-gravity problem. We describe in greater detail approaches which have not been covered in recent ``Quantum Gravity Phenomenology'' reviews. In particular, we outline a new framework for describing Lorentz invariance violation in the Maxwell sector. We also discuss the general strategy on the experimental side as well as on the theoretical side for a search for quantum gravity effects. The role of test theories, kinematical and dymamical, in this general context is emphasized. The present status of controlled laboratory experiments is described, and we also summarize some key results obtained on the basis of astrophysical observations.
\end{abstract}




\section{Introduction and preliminary remarks}

\subsection{The search for quantum gravity}

Our present description of the laws of physics may be characterized as obtained from two types of constituents. The first type of constituent are theoretical frameworks which  apply to {\it all} physical phenomena at {\it any} instant. These ``universal''  or ``frame''  theories are Quantum Theory (all matter is of microscopic origin), Special and General Relativity; SR and GR, (all kinds of matter locally have to obey the principles of Lorentz symmetry and behave in the same way in
gravitational fields), and statistical mechanics which is a method to deal with all kinds of systems for a large number of degrees of freedom. The second type of constituent is nonuniversal and pertains to the description of the four presently-known interactions: the electromagnetic, the weak, the strong and the gravitational. The first three interactions are all described within a single formalism, in terms of a gauge theory. So far only gravity has not been successfully included into that scheme. One reason for that might be that gravity appears on both sides: it is an interaction but it is at the same time also a universal theory. Universal theories like relativity and gravity are geometric in origin and do not rely on the particular physical system under consideration, whereas a description in terms of a particular interaction heavily makes use of the particular particle content. Therefore, gravity plays a distinguished role which may be the reason for the difficulty encountered in attempting to unify the other interactions with gravity and attempting to quantize gravity.

\renewcommand{\arraystretch}{1.1}
\tabcolsep10pt
\begin{center}
\begin{tabular}{|p{5.5cm}p{5.5cm}|}\hline
\rowcolor[gray]{0.9}{\bfseries Frame theories} & {\bfseries Interactions} \\ \hline
Quantum theory & Electrodynamics \\
Special Relativity & Gravity \\
General Relativity & Weak interaction \\
Statistical mechanics & Strong interaction \\ \hline
\rowcolor[gray]{0.9}{\bfseries Problem} & {\bfseries Wish} \\ \hline
Incompatibility of quantum theory and General Relativity & Unification of all interactions \\ \hline
\end{tabular}
\end{center}

The most pressing problem for present-day
theoretical physics is the unification of quantum theory
with gravity, the so-called ``quantum-gravity problem''.
The standard scheme of quantization has been proven to lead
to inconsistencies when applied to gravity.
In particular, the emerging theory would not be perturbatively
renormalizable. Various ideas have been explored in alternative to the standard
quantization procedure, and some of the most popular approaches
are based on string theory, canonical/loop quantum gravity, or non--commutative geometry. It is hoped that these quantum-gravity approaches may also provide a scheme for the unification of the four interactions, something that string theory already comes very close to doing.

A general feature of all these quantum-gravity scenarios is the appearance of new effects, often mediated by new fields, and in particular some of these effects appear to require a modification of some of the fundamental principles which SR and GR are based on.

\subsection{Implications of a new quantum gravity theory}

\begin{figure}[t]
\psset{unit=0.7cm}
\begin{center}
\begin{pspicture}(-9,-5)(5,5)
\pscircle[linecolor=light-blue,linewidth=10pt](0,0){4}
\rput(-4,0){\rnode{s}{
\ShowCX{white}{0.6cm}{s}}
}
\rput(-2.49396,3.12733){\rnode{m}{
\ShowCX{white}{0.6cm}{m}}
}
\rput(0.890084,3.89971){\rnode{A}{
\ShowCX{white}{0.6cm}{A}}
}
\rput(3.60388,1.73553){\rnode{mol}{
\ShowCX{white}{0.6cm}{mol}}
}
\rput(3.60388,-1.73553){\rnode{cd}{
\ShowCX{white}{0.6cm}{cd}}
}
\rput(0.890084,-3.89971){\rnode{K}{
\ShowCX{white}{0.6cm}{K}}
}
\rput(-2.49396,-3.12733){\rnode{kg}{
\ShowCX{white}{0.6cm}{kg}}
}
\rput(-9,-1.6){\rnode{UGR}{\txt\footnotesize{GR ensures \\ uniqueness of \\ timekeeping in \\ gravitational fields}}}
\rput(-9,1.6){\rnode{SR1}{\txt\footnotesize{SR ensures \\ uniqueness of \\ timekeeping in \\ moving frames }}}
\rput(-6.7,-3.8){\rnode{M}{\txt\footnotesize{GR ensures \\ uniqueness of \\ definition of mass}}}
%
\ncline[arrowscale=2 2]{->}{s}{m}\naput{\txt\footnotesize{SR: Constancy of \\ speed of light}}
\ncline[arrowscale=2 2]{->}{s}{A}
\ncline[arrowscale=2 2]{->}{s}{cd}
\ncline[arrowscale=2 2]{->}{m}{A}
\ncline[arrowscale=2 2]{->}{m}{cd}
\ncline[arrowscale=2 2]{->}{kg}{A}
\ncline[arrowscale=2 2]{->}{kg}{mol}
\ncline[arrowscale=2 2]{->}{kg}{cd}
\ncline[arrowscale=2 2]{->}{UGR}{s}
\ncline[arrowscale=2 2]{->}{SR1}{s}
\ncline[arrowscale=2 2]{->}{SR}{m}
\ncline[arrowscale=2 2]{->}{M}{kg}
\rput(-5,-1){$3\cdot 10^{-15}$}
\rput(-3.8,3.3){$10^{-12}$}
\rput(2.7,4){$4\cdot 10^{-8}$}
\rput(5.4,1.8){$8\cdot 10^{-8}$}
\rput(5,-1.8){$10^{-4}$}
\rput(2.7,-4){$3\cdot 10^{-7}$}
\rput(-2.5,-4.5){\txt\footnotesize{unknown ageing \\ of prototype}}
\end{pspicture}
\end{center}
\caption{The SI--units (s = second, m = meter,
A = Ampere, mol = mole, cd = candela,
K = Kelvin, kg = kilogram) and their interdependences \cite{Quinn95}.
The numbers indicate the stability of the corresponding unit.
The uniqueness of the transport of the definition of the second
and of the meter depends on the validity of SR and GR. A replacement
of the mechanical definition of the Ampere or other quantities also
requires conventional quantum theory and Maxwell theory and thus the
validity of the SR and GR. All units, but the Kelvin, are thus influenced
by relativity. \label{Fig:metrology}}
\end{figure}
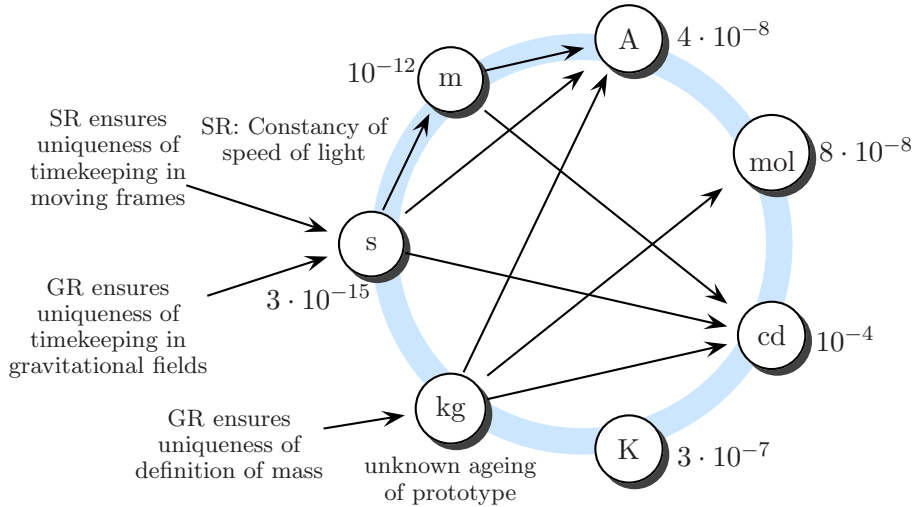

The possibility of solving the quantum-gravity problem is often described as intellectually exciting but of mere academic interest. However, the fact that a significant sample of quantum-gravity approaches appears to lead to modifications of some aspects of SR and GR suggests that the implications may go well beyond the academic interest. This point applies in particular to the area of modern metrology, the definition, preparation and transport of physical units. Since various atomic clocks on Earth are located at different height and geographical positions, it is clear that the uniqueness of the definition of time in terms of the TAI, the international atomic time, relies on the validity of SR and GR, see Fig.\ref{Fig:metrology}.

\begin{figure}[t]
\psset{xunit=1.2cm,yunit=0.8cm}
\begin{center}
\begin{pspicture}(-4,-3)(4,5)
\rput(0,4){\rnode{V}{
\ShowCX{white}{0.6cm}{V}}
}
\rput(-3.464,-2){\rnode{ohm}{
\ShowCX{white}{0.6cm}{$\Omega$}}
}
\rput(3.464,-2){\rnode{A}{
\ShowCX{white}{0.6cm}{A}}
}
\rput(0,2){\rnode{qV}{
\ShowCX{white}{0.5cm}{$\hbar/e$}}
}
\rput(-1.7321,-1){\rnode{qohm}{
\ShowCX{white}{0.5cm}{$\hbar/e^2$}}
}
\rput(1.7321,-1){\rnode{qA}{
\ShowCX{white}{0.5cm}{$e$}}
}
\rput(0,0){\txt\footnotesize{Validity \\ of Maxwell \\ and Schr\"odinger \\ equations}}
\ncline[arrowscale=2 2]{-}{V}{ohm}
\ncline[arrowscale=2 2]{-}{V}{A}
\ncline[arrowscale=2 2]{-}{V}{qV}
\ncline[arrowscale=2 2]{-}{ohm}{A}
\ncline[arrowscale=2 2]{-}{ohm}{qohm}
\ncline[arrowscale=2 2]{-}{A}{qA}
\ncline[arrowscale=2 2]{-}{qV}{qohm}
\ncline[arrowscale=2 2]{-}{qV}{qA}
\ncline[arrowscale=2 2]{-}{qohm}{qA}
\end{pspicture}
\end{center}
\caption{Electrical units of the voltage V (Volt), electrical resistance $\Omega$ (Ohm), and electrical current A (Ampere) based on distinguished quantum effects. The experimental effects are interpreted in terms of the ordinary Maxwell and Schr\"odinger equation. \label{Fig:ohmvoltampere}}
\end{figure}
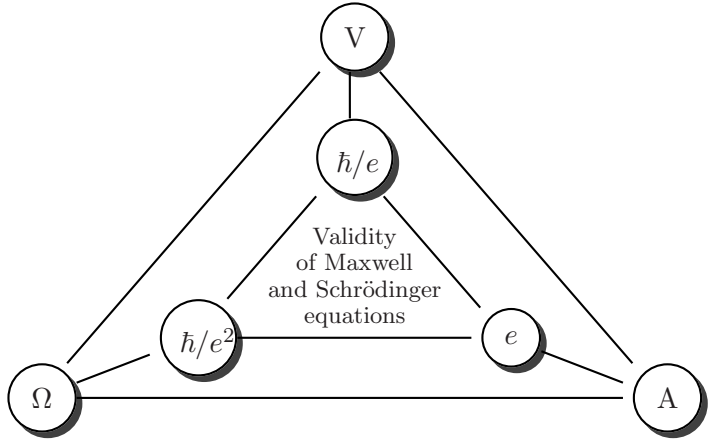

Another key objective of modern metrology is to base
all units on distinguished quantum effects.
The reason for that is the precision achieved
and the universal reproducibility which is based
on the uniqueness of quantum mechanics.
Beside the second and the meter, where
this already has been done, one can base
the unit of the electrical resistance,
the Ohm, on the quantum Hall effect
using $R_H = K/n$ where $n \in \mathbb{N}$
and $K = h/e^2$ is the von--Klitzing constant,
and the unit of the electrical voltage on the
Josephson effect via $U_J = n \nu/K_J$
where $K_J = 2e/h$ is the Josephson constant
and $\nu$ a given frequency, see Fig.\ref{Fig:ohmvoltampere}.
These quantum definitions of units heavily rely
on the validity of basic dynamical equations
like the Maxwell, the Schr\"odinger, and the
Dirac equation. If one of these equations was
to be modified then some definitions of units
would also be affected. Therefore, {\it any high precision test of
Lorentz invariance is also a test of the
modern metrological scheme}.

Moreover, if it turns out that the Maxwell and Dirac equations have to be modified then this may complicate future high precision navigation.

\subsection{The magnitude of quantum gravity effects}

Since the typical laboratory energies are of
the order of 1 eV and the characteristic quantum-gravity
energy scale is expected to be of the order
of the Planck energy which is about $10^{28}\;\hbox{eV}$,
the quantum-gravity effects in laboratory experiments are
likely to come in at the order of $10^{-28}$ (or lower)
which appears to be well beyond the reach of laboratory experiments,
in spite of the many high-precision devices which
are becoming available for searching for new effects.
As stressed in several recent ``quantum gravity phenomenology''
reviews~\cite{Amelino-Camelia00a,Sarkar02,JacobsonLiberatiMattingly04,Amelino-Camelia04} this suggests that the relevant phenomenology should
rely on contexts of interest in astrophysics, rather than use
controlled laboratory experiments.
However, there may nevertheless be an important role
for laboratory experiments in this phenomenology,
especially considering the following observations:
\begin{itemize}
\item The arguments that suggest that the characteristic quantum-gravity energy scale should be of
order $10^{28}\;\hbox{eV}$ must at present be viewed as inconclusive. One really needs the correct theory, which is still not established, in order to reliably estimate this scale. It may well be that the characteristic quantum-gravity energy scale is actually much lower than $10^{28}\;\hbox{eV}$. For example, in scenarios with ``large extra dimensions'' the quantum-gravity effects, including deviations from Newton's law, would be accessible at much lower energies
(see e.g. \cite{Antoniadis03}). Similarly, in
some string-theory-motivated ``dilaton scenarios''~\cite{DamourPiazzaVeneziano02,DamourPiazzaVeneziano02a} the Universality of Free Fall (UFF) would be violated already at the $10^{-13}$ level, and the PPN parameter $\gamma$, which in ordinary Einsteinian gravity is exactly 1, might be different from unity by up to $10^{-5}$.

\item  Even within the assumption that the quantum-gravity effects actually do originate at the Planck scale, one can find some (however rare) cases in which the small $10^{-28}$ effect is
effectively ``amplified''~\cite{Amelino-Camelia00a}
by some large ordinary-physics number that characterizes the laboratory setup.
For example, there has been considerable work on the possibility that quantum-gravity effects might significantly affect some properties of the neutral-kaon
system~\cite{Kostelecky99,Mavromatos04} and in those analyses the effects are amplified by the fact that the neutral-kaon system hosts the peculiarly small mass difference between long-lived and the short-lived kaons $M_{L,S}/|M_L - M_S| \sim 10^{14}$.

Similarly, in studies of quantum-gravity-induced interferometric
noise~\cite{AmelinoCamelia00} one can exploit
the fact that some scenarios predict noise that increases as the frequency of analysis decreases, and therefore, searching for fundamental noises in high precision long--term stable devices (like optical resonators) may give new access to this domain of quantum gravity
effects \cite{Schilleretal04}.

In an analogous way, the sensitivity of some clock--comparison experiments is amplified by a factor of $h \nu/m_p c^2 \sim 10^{18}$ where $\nu$ is the clock frequency used and $m_p$ the proton mass, see Sec.\ref{Sec:ClockComparisonExp}.

\item In some rare cases the laboratory experiment actually does have the required $10^{-28}$ sensitivity, as illustrated by the analysis reported
in Ref.~\cite{AmelinoCameliaLaemmerzahl04}, which studies the implications of quantum-gravity-induced wave dispersion for the most advanced modern interferometers (the interferometers whose primary intended use is the search of classical gravity waves).

\item The fact that low-energy data suggest that
the electroweak and strong interactions
unify at the GUT energy scale of $\sim 10^{16}\;\hbox{GeV}$ may
encourage us to conjecture that also the gravitational interaction
would be of the same strength as the other interactions at that
scale. This would mean that the characteristic energy scale
of quantum gravity should be some
three orders of magnitude smaller than the Planck energy
(see e.g. \cite{Duff98}).

\end{itemize}

It appears therefore that we should complement the astrophysics searches with a wide program of laboratory/controlled experiments searching for the effects predicted by some quantum-gravity approaches. Although in most cases the potential sensitivity of astrophysics searches will be higher, the laboratory experiments have the advantage of providing access to a much larger variety of potential phenomena, and for the development of the relevant phenomenology it would be an extraordinary achievement to be able to analyze a new effect within
a controlled/repeatable laboratory experiment with the possibility of a systematic modification of boundary and initial conditions, rather than relying on the ``one chance'' observations available in astrophysics.

\subsection{Main quantum gravity schemes}

As mentioned, the most popular approaches to the quantum-gravity problem are based on string theory, canonical/loop quantum gravity, or non--commutative geometry, see, e.g., \cite{Kiefer04}.
For different reasons the development of phenomenology for each of these three ideas is still at a rather early stage of development. But several hints for experimentalists have already
emerged.

\begin{itemize}
\item In canonical/loop quantum gravity the key difficulty is the fact that the techniques for obtaining the classical limit of the theory have not yet been developed.
Since our phenomenology will usually be structured as
a search of corrections to the classical effects, this is a
very serious issue. However, several
authors~\cite{GambiniPullin99,AlfaroMoralesTecotlUrrutia00,AlfaroMoralesTecotlUrrutia02,AlfaroMoralesTecotlUrrutia02a}, guided by the intuition from working with some candidate quasiclassical states, made analyses that led to the expectation that Lorentz symmetry should be broken in Loop Quantum Gravity, and as a result the Maxwell and Dirac equations should include extra terms of higher derivatives. But clearly these violations from Lorentz symmetry still cannot be viewed
as a ``prediction'' of Loop Quantum Gravity because of the heuristic nature of the underlying arguments, and indeed there are some authors (see, {\it e.g.} Ref.~\cite{KozamehParisi04})
who have presented arguments in favour of exact Lorentz symmetry for Loop Quantum Gravity.

\item String theory schemes in any dimensions always predict new fields which couple in different ways to the various matter sectors. As a result a large variety of effects are {\it allowed} by string theory, including effects leading to violation of Lorentz invariance, violation of the UFF and violation of the Universality of the Gravitational Redshift (UGR) is encountered. From the viewpoint of the phenomenologist a disappointing aspect of string theory is that it appears that it cannot be falsified on the basis of these effects. String theory may predict many new low-energy effects, but it can also be easily tuned to avoid all of them.
But of course it is still interesting to look for these new effects, and indeed a
rich phenomenology is being developed. In particular, there is considerable work on a string-theory-motivated dilaton scenario \cite{DamourPiazzaVeneziano02} with deviations from the UFF at the order $10^{-13}$ and deviations of the PPN parameters $\gamma$ and $\beta$ at the order $10^{-5}$ and $10^{-9}$, respectively. Another large phenomenological effort is being devoted
to a string-theory-motivated general framework, the Standard Model Extension (SME) \cite{ColladayKostelecky97}, for the description of violations of Lorentz invariance that are
codified in power-counting-renormalizable terms.

\item Much of the quantum-gravity work based on noncommutative geometry has focused on the hypothesis that the correct quantum gravity should have as flat-spacetime limit a noncommutative version of Minkowski spacetime, with spacetime coordinate noncommutativity
of the type $[{x}_\mu,{x}_\nu]= i\theta_{\mu\nu}
+ i \zeta_{\mu,\nu}^{\sigma} {x}_\sigma $ for some appropriate
choice of the coordinate-independent $\theta_{\mu\nu}$
and $\zeta_{\mu\nu}^{\sigma}$. In a noncommutative spacetime there is an absolute limitation
on the localization of a spacetime point (event),
and as a result the relevant theories are inevitably nonlocal.
In some frameworks it is possible to give an effective commutative-spacetime
description of this nonlocality, in which in particular -- in terms of partial differential equations -- higher order derivatives may occur. Another rather generic feature
of these noncommutative versions of Minkowski spacetime
is the emergence of anomalous dispersion relations.
The simplest and most studied example of
such an anomalous dispersion relation is
\begin{equation}
m^2 \simeq E^2 - \vec{p}^2 + \eta \vec{p}^2 {E \over E_{P}}
+ O\left({E^{4} \over E^{2}_{P}}\right)
\, , \label{ModDispersion}
\end{equation}
where $E_{\rm P}$ is the Planck scale  and $\eta$ is a numerical
factor (expected to be of order 1)
describing the strength of the quantum gravity modification
($E_{\rm QG} \equiv E_{\rm P}/\eta$ is the characteristric scale
of the modification of the dispersion relation).
\end{itemize}

\renewcommand{\arraystretch}{1.1}
\tabcolsep5pt
\begin{center}
{\footnotesize
\begin{tabular}{|p{0.2\textwidth}|p{0.2\textwidth}|p{0.2\textwidth}|p{0.2\textwidth}|} \hline
& \txt{string theory} & \txt{canonical/loop \\ quantum gravity} & \txt{non--commutative \\ geometry} \\ \hline
geometrical characteristics & starts from fixed (flat or deSitter) background & dynamical & starts from fixed background \\
particle characteristics & depends on particle model & model independent & model independent \\ \hline
violations of & Lorentz invariance & Lorentz invariance? (may depend on boundary conditions) & Lorentz invariance \\
& Universality of Free Fall & Universality of Free Fall? (not yet calculated) & Universality of Free Fall? (not yet calculated) \\ \hline
\end{tabular}}
\end{center}

\subsection{Methods of phenomenological generalizations of dynamical equations}

Since it was not covered extensively in recent quantum-gravity-phenomenology
reviews~\cite{Amelino-Camelia00a,Sarkar02,JacobsonLiberatiMattingly04,Amelino-Camelia04}
we will here devote particular attention to the study of phenomenological generalizations of dynamical equations like the Dirac or the Maxwell equations. As always in  quantum gravity phenomenology, these studies must be structured in terms of some test theories, bridging the gap between the rich (but often untreatable) formalisms used in the study of the full quantum gravity problem and the language which is appropriate to describe experiments.
The structure of these test theories should be obtained by calculating some low energy approximation of the full quantum gravity scenario. This has been attempted for the ``Liouville String'' scenario \cite{EllisMavromatosNanopoulos00,Ellisetal00}
as well as for the loop quantum gravity
approach \cite{GambiniPullin99,AlfaroMoralesTecotlUrrutia00,AlfaroMoralesTecotlUrrutia02,AlfaroMoralesTecotlUrrutia02a}. But since various technical issues remain to be understood about the relevant approximation schemes, the phenomenology is being developed on the basis of rather general test theories describing modifications of the Dirac and the Maxwell equations.

One approach, which is being pursued mostly as starting point of the SME \cite{ColemanGlashow97,ColladayKostelecky97,ColladayKostelecky98},
adopts some general Lagrangian which is still quadratic in the field
strengths or in the fermionic fields and requires further building
principles like conservation of energy-momentum, Lorentz-covariance,
conventional quantization, Hermiticity, microcausality, positivity
of energy, gauge invariance, and power-counting renormalizability.
The parameters of the SME are just additional interactions with
constant fields\footnote{This kind of generalizations in the photonic
sector have been introduced and discussed earlier by Ni \cite{Ni77}
and Haugan and Kauffmann \cite{HauganKauffmann95}.}. The main advantages of this approach are the mathematical consistency and physical interpretability of the new theory in conventional terms and
the fact that it provides the most conservative modification of
established theories. But of course this may also turn into a disadvantage, if it eventually
turns out that the correct quantum gravity requires more novel
features, such as effects that are not described by
power-counting-renormalizable terms. And indeed some
authors have chosen to look beyond the SME setup,
considering Planck-scale-suppressed effects which
are in fact not described by power-counting-renormalizable
terms (see, {\it e.g.},
Refs.~\cite{Amelino-Cameliaetal98,GambiniPullin99,MyersPospelov03}).

Of course, it is also possible to renounce to the assumption
of a Lagrangian generating the dynamical equations,
and in fact there is a rich phenomenology being
developed introducing the generalizations directly
at the level of the dynamical equations. This is of course more general than the Lagrangian
approach (see \cite{Laemmerzahl98,LaemmerzahlMaciasMueller05}
for examples dealing with generalized Dirac and Maxwell equation).
In the case of the Dirac equation \cite{Laemmerzahl98} one is led
to effects like the violation of the UFF,
of the UGR and of Local
Lorentz Invariance (the effects violating Local Lorentz Invariance
have been also obtained later on in \cite{ColladayKostelecky98}).
More effects than in the SME are encountered for the generalized
Maxwell equation discussed in \cite{LaemmerzahlMaciasMueller05}.
Charge conservation, which automatically comes out from the Lagragian
approach, can be violated in models generalizing the field
equations \cite{LaemmerzahlMaciasMueller05}. In addition,
in \cite{LaemmerzahlMaciasMueller05} also more Lorentz
Invariance violating parameters than in the SME have been found.
However, particular care must be taken for the mathematical
consistency of the formalism if one ''by hand'' generalizes
the dynamical equations. This will be automatically secured
if one employs a constructive axiomatic scheme in order to
derive equations, as shown for example in
Refs.~\cite{La90,AL93,AHL92} within a
derivation of the generalized Dirac equation in terms of
fundamental properties of the dynamics of fields.

\subsection{Kinematical test theories for Lorentz invariance}\label{Sec:KinematicalTestTheories}

Instead of modifying the dynamical equations, some authors have preferred
to introduce new-physics effects at the level of kinematics. In particular,
there is significant work on kinematical test theories for Lorentz invariance.
Later on in these notes we will comment on some recent proposals of this type
which have been motivated rather directly by some approaches to the
quantum-gravity problem. We here want to comment on a predecessor,
the Robertson--Mansouri--Sexl (RMS) theory, which was introduced
by Robertson \cite{Robertson49} and Mansouri and Sexl \cite{MansouriSexl77},
where a modification of Lorentz--transformations are the basis for, e.g.,
anomalous effects of light propagation what can be tested. Compared with
the dynamical approach the kinematical approach is more powerful since it
is independent of the specific particle model under consideration:
it discusses the transformation properties of observed quantities
under transformations between inertial systems. One may however
be uncomfortable with the fact that this RMS model assumes that
in one frame light propagates isotropic. This requires to single
out a preferred frame, which is usually taken to be the frame
given by the cosmological radiation background.
But this preferred role for the cosmological
radiation background, which is after all a
classical-physics feature, is not necessarily compelling.
And one might ask how this programme should proceed if it
turns out that, for example, there is also a stochastic
gravitational wave background, which selects a different
frame with respect to the one natural for the radiation background.
Should one then make a choice? This choice would influence significantly
the interpretation of the experiments. Some authors are also uncomfortable
with the RMS scheme because of its lack of generality with respect to
certain issues. In particular, it is not obvious that one should assume
that in the preferred frame light propagates isotropically, something
that might be violated in a Finslerian space--time,
and the RMS scheme also does not allow a description of possible violations
of Lorentz invariance leading to birefringence effects.

\subsection{The scheme of exploring new physics with generalized Maxwell and Dirac equations}

As mentioned in this notes we will devote particular attention to the possibility of exploring
new physics with generalized Maxwell and Dirac equations.
There is a long history in the discussion of violations of Lorentz invariance
and of the UFF (also called Weak Equivalence Principle) due to some modifications of the Maxwell and Dirac equations.

Very early considerations are based on Mach's principle and led to considerations of a hypothetical anomalous inertial mass term in the Schr\"odinger equation \cite{CocconiSalpeter58} which subsequently has been tested with very high precision by Hughes \cite{Hughesetal60} and Drever \cite{Drever61}, see also Sec.\ref{Sec:ClockComparisonExp}. Later discussions have been performed by Liebscher and Bleyer \cite{Liebscher85,BleyerLiebscher95}. The most general Dirac equation based on basic properties like unique evolution, superposition principle, finite propagation speed and probability conservation has been discussed in \cite{AL93}. Based on this, a generalized Pauli equation has been
derived \cite{Laemmerzahl98} which is most appropriate to confront this generalized theory with experiments. Since the spin--$\frac{1}{2}$--sector of the SME is
contained in this general scheme, also there a generalized Pauli equation can be derived \cite{KosteleckyLane99}.

Modifications of the Maxwell equations in view of a discussion of violations of Lorentz invariance and of the UFF has been first introduced by Ni \cite{Ni73,Ni77} and Haugan and Kauffmann \cite{HauganKauffmann95}. The same sort of modifications can be found in the SME \cite{ColladayKostelecky98,KosteleckyMewes02}. More general modifications which also include charge non--conservation have been introduced
in \cite{LaemmerzahlMaciasMueller05}.

For exploring new physics in these frameworks, at first one discusses and explores isolated effects for the Dirac and the Maxwell equation.
These are propagation phenomena of electromagnetic and matter waves and, in the case of electrodynamics, the static solutions for a point--like charge and magnetic moment. In a next step, combined effects which appear in electromagnetically bound systems have to be calculated, see, e.g., Sec.\ref{Sec:MacroscopicMatterEffects}. These combined effects govern the physics of atoms,  molecules, and solids which serve as realizations of clocks, and of rulers.

\begin{sidewaysfigure}
{\footnotesize
\begin{equation*}
\begin{xy} 0;<0.75cm,0cm>:
,(-5,13.3)*{\ShowXl{light-blue}{6.5cm}{\textbf{for quantum fields} \\ [-1ex]
\rule{\linewidth}{1pt} \\
$\bullet$ deterministic evolution \hfill\\
$\bullet$ superposition principle \\
$\bullet$ finite propagation speed \\
$\bullet$ probability conservation}}="AD"
,(-13,13.3)*{\begin{minipage}[t]{4cm}{\footnotesize
                 $\bullet$ H--atom \\
                 $\bullet$ solid \\
                 $\bullet$ interferometry \\
                 $\bullet$ neutrino propagation}\end{minipage}}="1Ex"
,(13,13.3)*{\begin{minipage}[t]{4cm}{\footnotesize
                 $\bullet$ birefringence \\
                 $\bullet$ dispersion \\
                 $\bullet$ damping}\end{minipage}}="TMaxwell1"
,(5,13.3)*{\ShowXl{light-blue}{6.5cm}{\textbf{for electromagnetic fields} \\ [-1ex]
\rule{\linewidth}{1pt} \\
$\bullet$ uniqueness of quantum mechanics \\
                               $\bullet$ deterministic evolution \\
                               $\bullet$ superposition principle \\
                               $\bullet$ finite propagation speed}}="AM"
,(-5,10.3)*{\ShowX{white}{6.5cm}{\begin{center} generalized Dirac equation  \\
                       $i \widetilde\gamma^\mu \partial_\mu \psi - M \psi = 0$
\end{center}}}="GD"
,(-13,10.3)*{\begin{minipage}[t]{4cm}{\footnotesize
                 $\phantom{\bullet}$ \\
                 $\bullet$ birefringence \\
                 $\bullet$ 2 mass shells \\
                 $\bullet$ spin--momentum--coupling \\
                 $\bullet$ neutrin propagation \\
                 $\bullet$ $g - 2$--experiments \\
                 $\bullet$ H--Atom}\end{minipage}}="2Ex"
,(13,10.3)*{\begin{minipage}[t]{4cm}{\footnotesize
                 $\bullet$ birefringence \\
                 $\bullet$ dispersion \\
                 $\bullet$ damping}\end{minipage}}="TMaxwell2"
,(-5,8)*{\ShowX{white}{6.5cm}{\begin{center} generalized
Pauli--equation \\
$\displaystyle i \partial_t \psi = \frac{1}{2m} \Bigl(\delta^{ij} +
\frac{\delta m_{\hbox{\scriptsize I}}^{ij}}{m}\Bigr) \partial_i \partial_j
\psi + \ldots$  \end{center}}}="GP"
,(5,9.8)*{\ShowX{white}{6.5cm}{\begin{center} generalized
Maxwell--equations \end{center}   \vspace*{-8mm}  \begin{eqnarray*}
\partial_{[\mu} F_{\nu\rho]} & = & 0 \\    \partial_\mu
(\lambda^{\mu\nu\rho\sigma} F_{\rho\sigma}) + \lambda^{\nu\rho\sigma}
F_{\rho\sigma} & = & 0 \end{eqnarray*}}}="GM"
,(0,4.9)*{\ShowX{white}{5.5cm}{\begin{center}bound
systems\end{center}\vspace*{0.8cm}}}="BS"
,(-1.7,4.6)*{\ShowX{white}{2cm}{\begin{center}\textcolor{white}{p}atoms\textcolor{white}{p}\end{center}}}="AT"
,(1.7,4.6)*{\ShowX{white}{2cm}{\begin{center}solids\end{center}}}="SD"
,(-5,2)*{\ShowX{white}{3cm}{c-o-m motion.: \\
atomic interferom.\textcolor{white}{y}}}="AI"
,(0,2)*{\ShowX{white}{3cm}{relative motion: \\ spectroscopy, clocks }}="SP"
,(5,2)*{\ShowX{white}{3cm}{\begin{center}\textcolor{white}{y}rulers\textcolor{white}{y}\end{center}}}="RL"
,(0,-0.4)*{\ShowX{light-red}{14cm}{\vspace*{1.3cm}\begin{center}Tests of
Special and General Relativity = Tests of the
EEP\end{center}}}="T"
,(-5,-0.1)*{\ShowX{white}{3cm}{\begin{center}UFF\end{center}}}="X2"
,(0,-0.1)*{\ShowX{white}{3cm}{\begin{center}redshift,
Doppler--effect\end{center}}}="X3"
,(5,-0.1)*{\ShowX{white}{3cm}{\begin{center}Michelson--Morley,
Kennedy--Thorndike\end{center}}}="X4"
\ar@{->} "AD";"GD"
\ar@{->} "AM";"GM"
\ar@{->} "GD";"GP"
\ar@{->} "GP";"BS"
\ar@{->} "GM";"BS"
\ar@{->} "AT";"AI"
\ar@{->} "AT";"SP"
\ar@{->} "SD";"RL"
\ar@{->} "AI";"X2"
\ar@{->} "SP";"X3"
\ar@{->} "RL";"X4"
\ar@{->} "AD";"1Ex"
\ar@{->} "GD";"2Ex"
\ar@{->} "AM";"TMaxwell1"
\ar@{->} "GM";"TMaxwell2"
\end{xy}
\end{equation*}}
\caption{Schematics of the implications of modified Dirac and Maxwell equations.}
\end{sidewaysfigure}
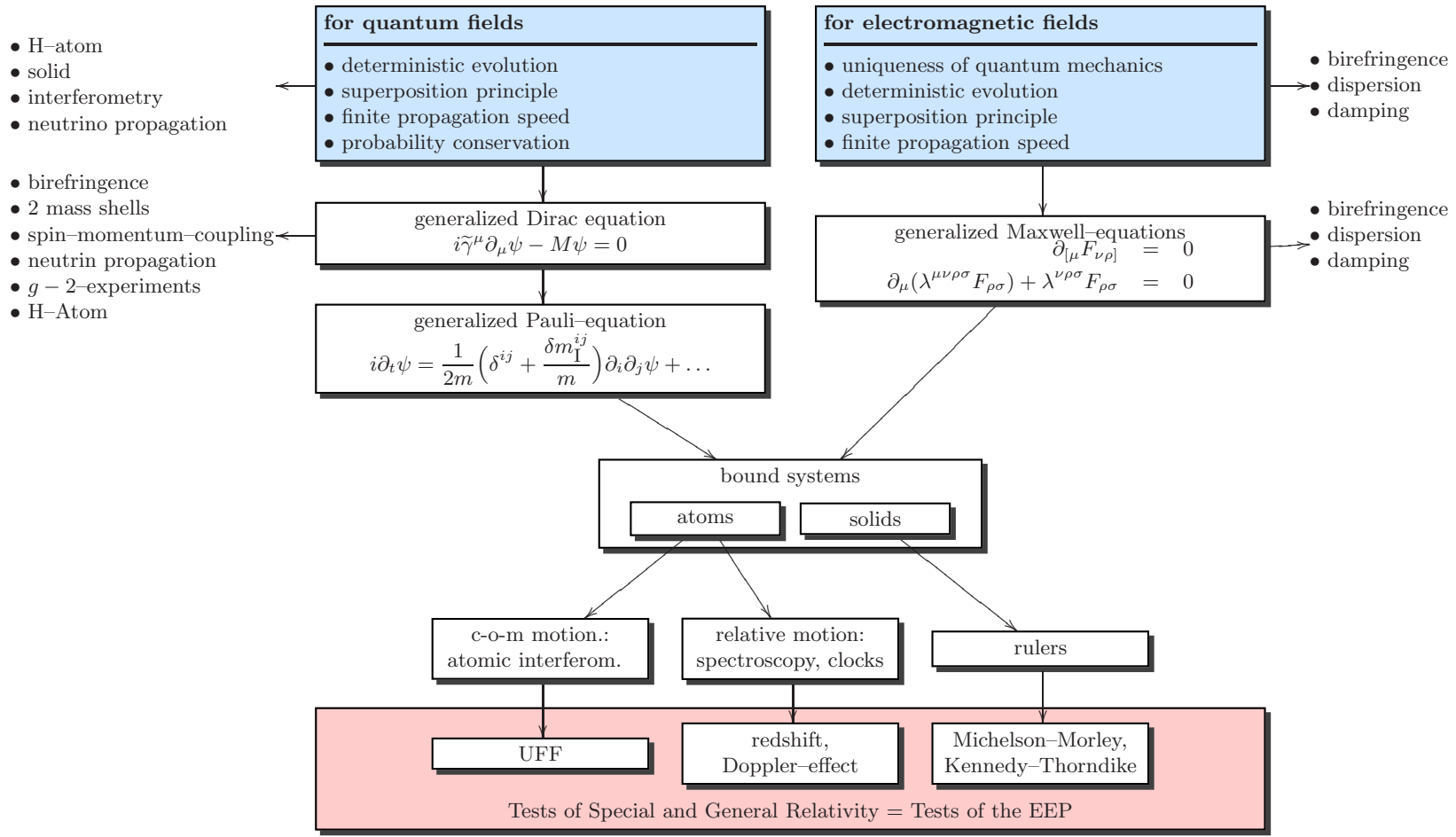

\subsection{What is a violation of Lorentz invariance?}

A key focus for these notes are studies of possible departures from Lorentz symmetry. For this phenomenology a key question is how we may interpret or how we may proceed if we are detecting a signal which appears to violate Lorentz invariance. After all the attempts at a conventional-physics explanation have failed, and it is therefore certain that one is dealing with new physics, it would be nice to have some criteria for establishing what constitutes a robust basis for accepting or recognizing a violation of Lorentz invariance.

Of course in science one never establishes how things ``are'', but one rather finds models that reproduce all observations. Therefore any interpretation will always be subject to further scrutiny. Still it is useful to keep in mind that some checks of consistency with the hypothesis of violation of Lorentz symmetry can be made. In particular, it is (at least in principle) relatively easy to distinguish between the hypothesis of a violation of Lorentz symmetry and the hypothesis of the discovery of a new interaction. If the effect signals a new interaction it should be possible to shield it (like the electromagnetic interaction), transform it away (like metrical gravity) and/or identify a cause of the effect. A violation of Lorentz invariance should instead be universal and it should not be possible to transform it away.

\subsection{Main directions in searches for new physics}

At present there are at least six main directions in the search for possible quantum gravity effects:
\begin{description}
\item[Search for orientation dependent effects.] The effects one is looking for depend on the orientation of the laboratory. Early experiments of this kind are Michelson--Morley (see Sec.\ref{SMEcavexp}) and Hughes--Drever experiments (see Sec.\ref{Sec:ClockComparisonExp}). For power-counting-renormalizable effects the results of all experiments of this kind can be related to the SME, which gives the most general parametrization of power-counting-renormalizable anisotropy effects in various basic equations. Presently, for photons the anisotropy is limited by $10^{-15}$ and for nuclear matter by $10^{-30}$. 

\item[Search for a violation of the Universality of Free Fall.] Within the string-theory-inspired dilaton scenarios there is a distinguished prediction of a violation of the UFF at the $10^{-13}$ level \cite{DamourPiazzaVeneziano02}. Presently, the UFF is confirmed at the level $5 \cdot 10^{-13}$.

Also anomalous spin couplings are considered (see, e.g., \cite{Laemmerzahl98b} for a short review), which are beyond the standard coupling to the gravitational field. Other works considering possible modifications of the UFF for various particles are \cite{Gasperini88,AdunasRodriguez-MillaAhluwalia01}.

\item[Search for a violation of the Universality of the Gravitational Redshift.] String theory
may also host deviations from the UGR \cite{Damour00}. In fact, it has been shown that violations and UFF and of UGR are strongly related \cite{DvaliZaldarragia02,Nordtvedt03}.

Presently, the UGR is confirmed by clock--clock comparison (atomic fountain clock vs. H--maser) at the $1.5 \cdot 10^{-5}$--level \cite{BauchWeyers02}.

\item[Search for a modified dispersion relation.] Searches of modifications of the dispersion relation of the form (\ref{ModDispersion}), and its generalizations, constitute another direction of experimental quantum gravity efforts. This type of effect is
beyond the SME, since it could not be described by power-counting-renormalizable terms.
A possible manifestation of the modified dispersion relations is an energy dependent propagation velocity which should lead to different time--of--arrivals of the same event on a distant star when looked at it in different frequency channels, see e.g. \cite{Schaefer99}. Another effect is a frequency dependent position of interference fringes \cite{AmelinoCameliaLaemmerzahl04}.

\item[Search for space--time fluctuations.] Search for fundamental space--time
fluctuations~\cite{AmelinoCamelia00} and associated fundamental decoherence effects in quantum systems~\cite{Percival95,PercivalStrunz96} have been also considered extensively in association with various classes of experiments~\cite{Ellisetal84,Schilleretal04}. There has also been work on the implications of space--time fluctuations for the spreading of signals, which causes a sharp signal to turn into a fuzzy signal upon relatively long propagation
times~\cite{Amelino-Camelia00a,YuFord99}.

\item[Search for departures from $CPT$ symmetry]
Since the $CPT$ theorem is based on locality and Lorentz covariance, it is not surprising that the quantum-gravity approaches that involve some nonlocality and/or departures from Lorentz symmetry provide motivation for searches of departures from $CPT$ symmetry.
Noteworthy limits have been obtained using some analyses of the neutral-kaon system (see Ref.~\cite{Ellisetal96} and references therein) and other limits are now being pursued in the context of neutrino physics~\cite{Mavromatos04}. Indeed, it has been shown that $CPT$ violation implies a violation of Lorentz invariance \cite{Greenberg02}.

\end{description}

\begin{figure}
\begin{center}
\begin{pspicture}(0,1)(15,5)
\rput(7.5,5){Search for quantum gravity effects}
\rput(1.5,4.15){\txt\footnotesize{Search for \\ anisotropies}}
\rput(1.5,2.3){\includegraphics[width=2.5cm]{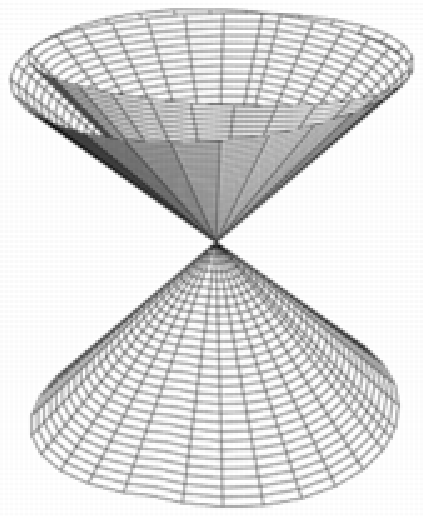}}
\rput(4.5,4){\txt\footnotesize{Search for \\ violations \\ of UFF}}
\pscircle[fillstyle=solid,fillcolor=black](4,2.7){0.3}
\pscircle[fillstyle=solid,fillcolor=yellow](5,2.7){0.3}
\psline{->}(4,2.5)(4,1.5)
\psline{->}(5,2.5)(5,1.5)
\rput(4,1.2){$g$}
\rput(5,1.2){$g$}
\rput(7.5,4){\txt\footnotesize{Search for \\ violations \\ of UGR}}
\rput(6.8,2){\includegraphics[width=1.4cm]{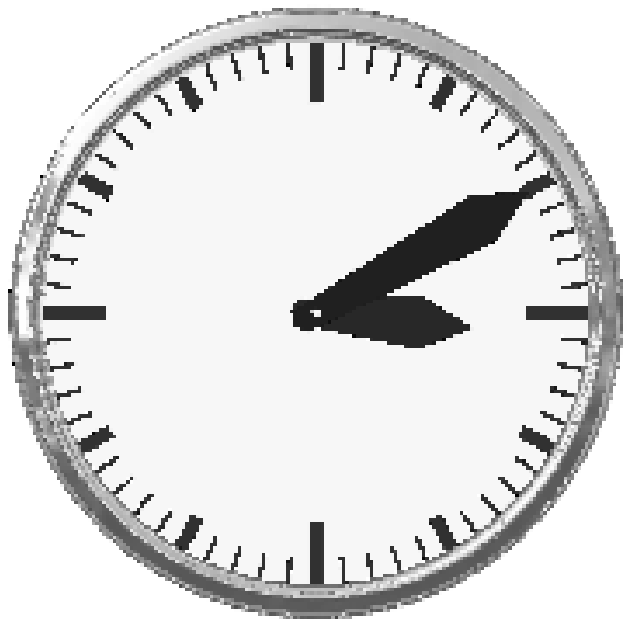}}
\rput(8.2,2){\includegraphics[width=1.4cm]{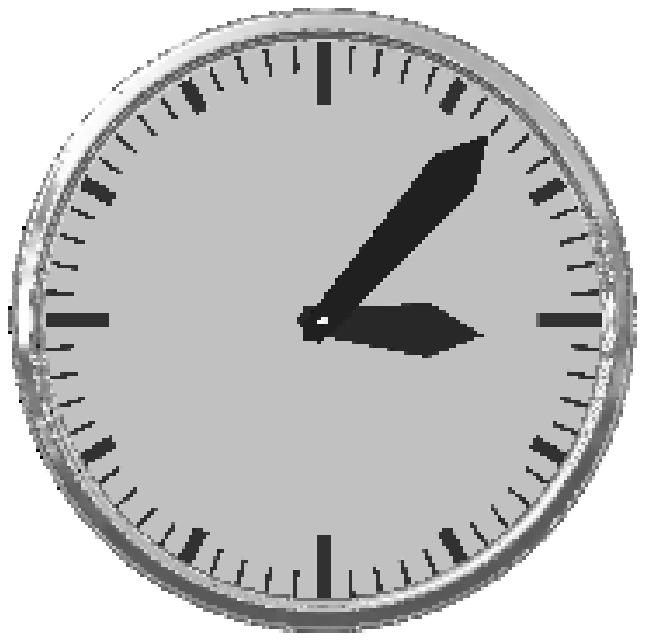}}
\rput(10.5,4){\txt\footnotesize{Search for \\ anomalous \\ dispersion}}
\rput(10.5,1.2){
\psline{->}(-1.2,0)(1.2,0)
\psline{->}(0,0)(0,1.5)
\rput(0.2,1.5){\footnotesize $\omega$}
\rput(1.2,0.2){\footnotesize $k$}
\psplot[linestyle=dashed]{0}{1.2}{x}
\psplot[linestyle=dashed]{-1.2}{0}{x -1 mul}
\psplot{0}{1.2}{x x dup mul 0.2 mul add}
\psplot{-1.2}{0}{x -1 mul x dup mul 0.2 mul add}
}
\rput(13.5,4){\txt\footnotesize{Search for \\ space--time \\ fluctuations}}
\rput(13.5,2){\includegraphics[width=3cm,height=2.5cm]{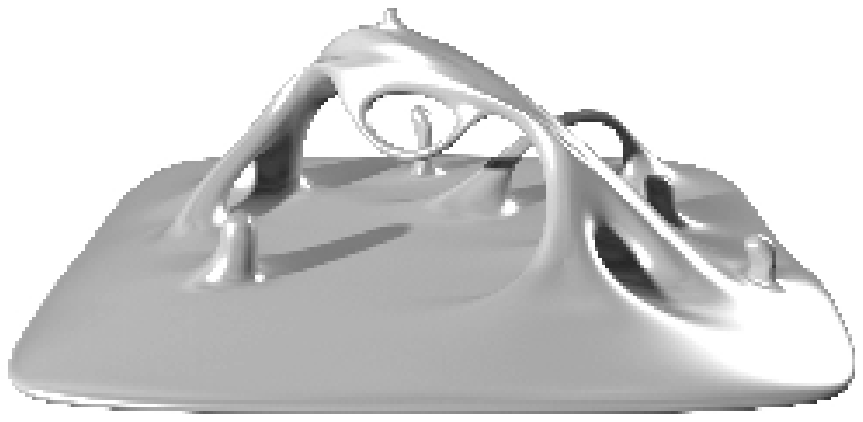}}
%
\rput(4.5,0.8){$\underbrace{\phantom{xxxxxxxxxxxxxxxxxxxxxxxxxxxxxxxxxxxxxxx}}$}
\rput(4.5,0.4){Einstein Equivalence Principle}
\end{pspicture}
\end{center}
\medskip
\caption{The main directions and characteristics of searches for
quantum gravity induced effects: anisotropy effects, violation
of UFF and UGR, modified dispersion and space--time fluctuations.
A search for violations of isotropy, UFF and UGR amounts
to a search of a violation of Einstein's Equivalence Principle.}
\end{figure}

\section{Quantum gravity predictions}

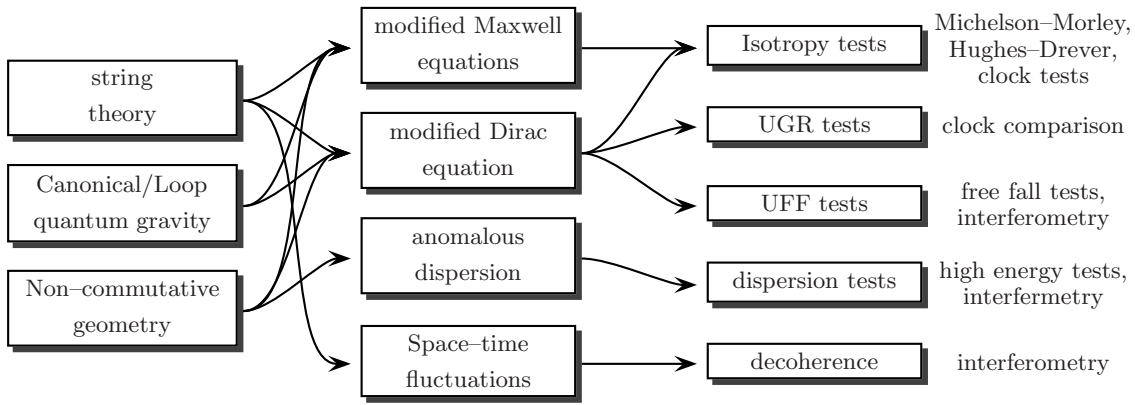
\begin{figure}[t]
\psset{xunit=1.15cm,yunit=0.7cm}
\begin{center}
\begin{pspicture}(-4,-4)(6,5)
\rput(-4,2){\rnode{ST}{
\ShowX{white}{2.8cm}{\footnotesize string \\ theory}}
}
\rput(-4,0){\rnode{LQG}{
\ShowX{white}{2.8cm}{\footnotesize Canonical/Loop \\ quantum gravity}}
}
\rput(-4,-2){\rnode{NCG}{
\ShowX{white}{2.8cm}{\footnotesize Non--commutative \\ geometry}}
}
\rput(0,3){\rnode{GM}{
\ShowX{white}{2.6cm}{\footnotesize modified Maxwell equations}}
}
\rput(0,1){\rnode{GD}{
\ShowX{white}{2.6cm}{\footnotesize modified Dirac equation}}
}
\rput(0,-1){\rnode{AD}{
\ShowX{white}{2.6cm}{\footnotesize anomalous dispersion}}
}
\rput(0,-3){\rnode{STF}{
\ShowX{white}{2.6cm}{\footnotesize Space--time fluctuations}}
}
\rput(4,3){\rnode{LI}{
\ShowX{white}{2.6cm}{\footnotesize Isotropy tests}}
}
\rput(4,1.5){\rnode{UGR}{
\ShowX{white}{2.6cm}{\footnotesize UGR tests}}
}
\rput(4,0){\rnode{UFF}{
\ShowX{white}{2.6cm}{\footnotesize UFF tests}}
}
\rput(4,-1.5){\rnode{Disp}{
\ShowX{white}{2.6cm}{\footnotesize dispersion tests}}
}
\rput(4,-3){\rnode{Dec}{
\ShowX{white}{2.6cm}{\footnotesize decoherence}}
}
\rput(6.5,3){\txt\footnotesize{Michelson--Morley, \\ Hughes--Drever, \\ clock tests}}
\rput(6.5,1.5){\txt\footnotesize{clock comparison}}
\rput(6.5,0){\txt\footnotesize{free fall tests, \\ interferometry}}
\rput(6.5,-1.5){\txt\footnotesize{high energy tests, \\ interfermetry}}
\rput(6.5,-3){\txt\footnotesize{interferometry}}
\nccurve[angleB=180,ncurv=0.5,arrowscale=2 2]{->}{ST}{GD}
\nccurve[angleB=180,ncurv=0.5,arrowscale=2 2]{->}{ST}{GM}
\nccurve[angleB=180,ncurv=0.5,arrowscale=2 2]{->}{ST}{STF}
\nccurve[angleB=180,ncurv=0.5,arrowscale=2 2]{->}{LQG}{GD}
\nccurve[angleB=180,ncurv=0.5,arrowscale=2 2]{->}{LQG}{GM}
\nccurve[angleB=180,ncurv=0.5,arrowscale=2 2]{->}{NCG}{GD}
\nccurve[angleB=180,ncurv=0.5,arrowscale=2 2]{->}{NCG}{GM}
\nccurve[angleB=180,ncurv=0.5,arrowscale=2 2]{->}{GD}{UFF}
\nccurve[angleB=180,ncurv=0.5,arrowscale=2 2]{->}{GD}{LI}
\nccurve[angleB=180,ncurv=0.5,arrowscale=2 2]{->}{GD}{UGR}
\nccurve[angleB=180,ncurv=0.5,arrowscale=2 2]{->}{GM}{LI}
\nccurve[angleB=180,ncurv=0.5,arrowscale=2 2]{->}{AD}{Disp}
\nccurve[angleB=180,ncurv=0.5,arrowscale=2 2]{->}{STF}{Dec}
\nccurve[angleB=180,ncurv=0.5,arrowscale=2 2]{->}{NCG}{AD}
\end{pspicture}
\end{center}
\vspace*{-8mm}
\caption{Quantum gravity, effective intermediate (test) theories and the experimental search.}
\end{figure}

\subsection{String theory}

The low-energy physics of String theory, as presently understood,
could be very rich, with the presence of new fields
and the possibility of a variety of new effects.
This may provide the basis for a large phenomenological
effort, even though one should keep in mind that the new effects
are not genuine ``predictions'' of String theory: String theory
could make room for these effects but it could equally well suppress
them all. It appears at present not possible to falsify
String theory on the basis of low-energy phenomenology, but
it would nonetheless be very exciting
if any of the new effects that String theory may host
was actually found.

\paragraph{Lorentz invariance}

A possible violation of Lorentz invariance, due to
spontaneous symmetry breaking, has been
considered \cite{KosteleckySamuel89a}.
Though requirements like renormalizability
and gauge invariance rule out spontaneous
symmetry breaking of the Lorentz group in
ordinary field theories it might happen for
string theories. Lorentz symmetry breaking
may arise from tensor--tensor--scalar
couplings that are allowed because strings
are extended objects which contain an
infinite number of particle modes.

\paragraph{Universality of Free Fall}

The UFF is the most basic principle underlying GR. It is responsible for the
possibility to geometrize the gravitational interaction
and, thus, for the present understanding of space and time.
However, in some string-inspired dilaton scenarios \cite{DamourPolyakov94,DamourPolyakov94a,DamourPiazzaVeneziano02,DamourPiazzaVeneziano02a}
is has been claimed that the UFF might be violated, in terms of the E\"otv\"os parameter,
at the $10^{-13}$ level. Similar predictions are made by Wetterich \cite{Wetterich03} from a quintessence model where the dynamics of a cosmological field introduced to explain the dark energy, yields a violation of the UFF at the $10^{-14}$ level. Also in the ``Liouville noncritical string'' scenario \cite{Ellisetal03} there are predictions for a violation of UFF due to
the interaction of particles with the space--time foam which, however, are too small to be of today's experimental interest.

\paragraph{Universality of Gravitational Redshift}

The same string-theory scalar field,
which may couple to different matter fields in a
different way thus leading to a violation of UFF,
also leads to the effect that clocks based on
different physical principles behave differently
in the gravitational field and, thus, violate the UGR. This
has been shown in \cite{Damour97,Damour00}.
The relation between violations of UFF and UGR
has been discussed in \cite{DvaliZaldarragia02,Nordtvedt03}.
Also quintessence
models predict a violation of UGR \cite{Wetterich88}.

\paragraph{Anomalous dispersion}

Since, in principle, the effective equations for the electromagnetic field and also for the Dirac field contain higher order derivatives, anomalous dispersion relations of the structure given in Eq.(\ref{ModDispersion}) could also come out from string theory \cite{ColladayKostelecky98}. However, in most cases they are neglected in the string-theory context since it is believed that they are less significant than the modifications, e.g., of the Dirac matrices in the case of the Dirac equation or of the constitutive relation in electrodynamics.

\paragraph{Space--time fluctuations and decoherence}

In the most popular ``critical superstring'' formulation of String Theory there has been so far no direct argument in favour of spacetime fluctuations and decoherence. However, in the ``Liouville noncritical String'' approach~\cite{EllisMavromatosNanopoulos92} space--time fluctuations and decoherence have been considered. For decoherence, possible implications for the neutral-kaon system have been investigated (see, {\it e.g.}, Ref.~\cite{GravanisMavromatos01} and references therein), while a direct investigation of spacetime fluctuations has been attempted by Percival and coworkers \cite{Percival95,PercivalStrunz96}.

\subsection{Loop quantum gravity}

As mentioned, in canonical/loop quantum gravity the key difficulty is
the fact that the techniques for obtaining the classical limit of the theory have not yet been developed. In a certain sense one has an otherwise attractive candidate for quantum gravity, which however has not been shown to actually contain classical gravity as a limit.
As a way to get some intuition for the type of effects that the theory might predict, when fully understood, several authors~\cite{GambiniPullin99,AlfaroMoralesTecotlUrrutia00,AlfaroMoralesTecotlUrrutia02,AlfaroMoralesTecotlUrrutia02a} have proposed to start with the exploration of the properties of some candidate quasiclassical states.

\paragraph{Lorentz invariance}

By introducing the so--called weave states, which are candidate
quasiclassical states of the space--time geometry, in
the range between pure quantum geometry and classical
geometry, one can motivate
effective equations for the propagation of spinors and of
electromagnetic
fields \cite{AlfaroMoralesTecotlUrrutia00,AlfaroMoralesTecotlUrrutia02,AlfaroMoralesTecotlUrrutia02a,Alfaroetal04}.
The resulting effective equations possess
terms which violate Lorentz invariance, and also terms of higher
order derivative. As mentioned there are some critiques on this approach,
notably by Kozameh \cite{KozamehParisi04} who claimed that the
breaking of LI came in by a particular choice of boundary conditions.
Recently, it was also observed~\cite{Amelino-CameliaSmolinStarodubtsev04} that a plausible
path toward the Loop-Quantum-Gravity classical limit should lead to
a situation in which Lorentz symmetry is neither preserved nor broken,
it would be deformed, in the sense we discuss more carefully later in
our remarks on noncommutative spacetimes.

\paragraph{Universality of Free Fall and Unversality of Gravitational Redshift}

To our knowledge there has been no analysis so far suggesting violations of the UFF and UGR in Loop Quantum Gravity. This is not surprising in light of the mentioned ``classical limit''
problem: one should first verify UFF and UGR in the classical limit and then
look for possible quantum corrections.

\paragraph{Anomalous dispersion}

A modification of the energy-momentum dispersion relation is rather typical (though not inevitable) when Lorentz symmetry is broken. The arguments \cite{AlfaroMoralesTecotlUrrutia00,AlfaroMoralesTecotlUrrutia02,AlfaroMoralesTecotlUrrutia02a,Alfaroetal04} that suggest breaking of Lorentz symmetry in Loop Quantum Gravity do indeed provide motivation for modifications of the dispersion relation, since they lead to modified Maxwell and Dirac equations with higher order derivatives, suggesting a modified dispersion relation of the structure (\ref{ModDispersion}).

\paragraph{Space--time fluctuations and decoherence}

There has not been much discussion of Space--time fluctuations and decoherence in Loop Quantum Gravity. However, a recent proposal~\cite{GambiniPortoPullin03} for a reformulation of Loop Quantum Gravity does lead to decoherence and the need to set up the analysis within a density-matrix framework.

\subsection{Non--commutative geometry}

The formalism of noncommutative geometry appears to have rather
wide (though not completely general) applicability in the study of the
quantum-gravity problem. In particular, a large number of studies
has been devoted to the hypothesis that the correct quantum gravity
might admit a regime which
could be based on a non-commutative version of Minkowski
spacetime. In some cases non-commutative spacetimes prove
useful at an effective-theory level (for example, in certain string theory
pictures~\cite{DouglasNekrasov01} spacetime noncommutativity provides
an effective theory description of the physics of strings in presence
of a corresponding external background), while other quantum-gravity
approaches (see, {\it e.g.}, Ref.~\cite{DoplicherFredenhagenRoberts94})
explore the possibility that a noncommutativity might be needed for the correct
fundamental description of spacetime.

The most studied noncommutative versions of Minkowski spacetime all fall
within the parametrization ($\mu=0,1,2,3$)
\begin{equation}
[{x}_\mu,{x}_\nu]= i\theta_{\mu\nu} + i \zeta_{\mu\nu}^{\sigma} {x}_\sigma ~,\label{general}
\end{equation}
with coordinate-independent $\theta_{\mu\nu}$ and $\zeta_{\mu\nu}^{\sigma}$.
The choice of $\theta_{\mu\nu}$ and $\zeta_{\mu\nu}^{\sigma}$ is
the key theoretical input and the aspect to be determined experimentally.

\paragraph{Lorentz invariance}

An intense research effort has been devoted to
the implications of noncommutativity for the classical
Poincar\'{e} symmetries of Minkowski spacetime.
This is in fact one of the few aspects of the relevant theories
which can be analyzed rather exhaustively, and it leads to some
ideas for experimental testing.
For the simplest noncommutative versions of Minkowski spacetime,
the canonical noncommutative spacetimes characterized by coordinate
noncommutativity of type
\begin{equation}
[{x}_\mu,{x}_\nu]= i\theta_{\mu\nu}\label{canonical}
\end{equation}
with coordinate-independent $\theta_{\mu\nu}$,
a full understanding has been matured, and in particular it has been
established that the Lorentz-sector symmetries are
broken~\cite{DouglasNekrasov01,MatusisSusskindToumbas00,AmelinoCameliaetal03}
(in a sense analogous to the popular mechanism of spontaneous
symmetry breaking)
by this type of noncommutativity.

At the next level of complexity~\cite{Madoreetal00},
the one of Lie-algebra noncommutative spacetimes
\begin{equation}\label{liealg}
[{x}_\mu,{x}_\nu]= i \zeta_{\mu\nu}^{\sigma} {x}_\sigma ~,
\end{equation}
the description of symmetries is in some cases more demanding
at the technical level.
The most studied of these Lie-algebra versions of Minkowski
spacetime is ``$\kappa$-Minkowski~\cite{MajidRuegg94,LukierskiRueggZakrzewski95}
($l,m = 1,2,3$)
\begin{equation}
\left[x_m,t\right] = {i \over \kappa} x_m ~,~~~~\left[x_m, x_l\right] = 0 ~.
\label{kmindef}
\end{equation}
$\kappa$-Minkowski is a Lie-algebra spacetime that clearly
enjoys classical space-rotation
symmetry; moreover, at least in
a Hopf-algebra sense (see, {\it e.g.},
Ref.~\cite{AmelinoCameliaArzano02,AgostiniAmelinoCameliaDAndrea03}),
 $\kappa$-Minkowski is invariant under ``noncommutative translations''.
 It is particularly noteworthy that in $\kappa$-Minkowski boost transformations
are necessarily
modified~\cite{AmelinoCameliaArzano02,AgostiniAmelinoCameliaDAndrea03,KowalskiGlikmanNowak02}.
A first hint of this comes from the necessity
of a deformed law of composition of momenta, encoded
in the so-called coproduct  (a standard structure for a Hopf algebra).
As a result in this spacetime one expects
departures from classical Lorentz symmetry
but of a type that does not lead to the
emergence of a preferred class of inertial
observers (the ``Relativity Principle'' is still upheld).
One usually refers to these cases as ``{\underline{deformations}}
of Lorentz symmetry''
(in alternative to the more familiar mechanisms that may lead
to {\underline{broken}} Lorentz symmetry).

\paragraph{Unversality of Free Fall and Universality of Gravitational Redshift}

Concerning the UFF and the UGR these approaches based on noncommutative versions of Minkowski spacetime are insufficient for the derivation of rigorous predictions, since indeed they focus on the Minkowski limit. They are motivated by the idea of some form of ``quantization'' of gravity, but at present they only incorporate this quantization in the noncommutativity of spacetime and gravitational effects are not yet described (although some attempts of generalization are under way).

\paragraph{Anomalous dispersion}

The fact that noncommutative versions of Minkowski spacetime typically
lead to departures from classical Lorentz symmetry (broken Lorentz symmetry in most cases,
and, as mentioned, ``deformed'' Lorentz symmetry in some special
cases like $\kappa$-Minkowski) in turn leads to the emergence of departures
from the special-relativistic energy-momentum (``dispersion'') relation.

For the cases affected by IR/UV (infrared/ultraviolet) mixing the phenomenology based on modified dispersion relations must be conducted cautiously, since the IR/UV mixing may lead to infrared singularities, which are still not fully understood (or at least there is still a lively debate on how to handle them in phenomenology). We do not to comment on this possible phenomenology in these notes, but refer to Refs.~\cite{Carrolletal01,AmelinoCameliaetal03,CarlsonCaroneLebed02}.

When there is no singularity in the infrared, and assuming that the onset of the new effects is characterized by the Planck length, one will typically find (and indeed one finds in the examples which have been analyzed) dispersion relations that
fall within the class
\begin{equation}
m^2 \simeq E^2 - \vec{\mbox{\boldmath$p$}}^2 + \eta {\mbox{\boldmath$p$}}^2 \left({E^n \over E^n_{P}}\right)
+ O\left({E^{n+3} \over E^{n+1}_{P}}\right) \, , \label{displeadbis}
\end{equation}
where $\eta$ and $n$ are model-dependent parameters which should be determined experimentally. Not only are $\eta$ and $n$ model dependent, but within a given model they may also take different values for different particles. In particular, in the analysis~\cite{MatusisSusskindToumbas00} of some noncommutative spacetimes one finds birefringence (different polarizations of light propagate at different speeds).
The fact that some noncommutative spacetimes host this particle dependence
suggests that they might also give rise to departures from the UFF and the UGR,
but, as mentioned, the relevant formalisms are still at
too early a stage of development for producing detailed predictions
concerning the UFF and the UGR.

\paragraph{Space--time fluctuations}

One of the original motivations for introducing the formalism of
noncommutative geometry was the one of formalizing the limitations
on the localization of spacetime points (events) at the quantum-gravity
(Planck-scale) level, which are suggested by a variety of arguments
combining general relativity and quantum mechanics
(see, {\it e.g.},
Refs.~\cite{Padmanabhan87,DoplicherFredenhagenRoberts94,Ahluwalia94,NgVanDam94,AmelinoCamelia94,AmelinoCamelia96,Garay95}).
Unfortunately the analysis of the relevant formalisms has not
yet been developed to the point of a phenomenologically useful
description of this localization limits.
The noncommutativity of the coordinates of course implies some
associated limits on the combined measurement of pairs
of coordinates, but phenomenology would need a
more advanced level of description, one in which one could for example
derive the implications of noncommutative geometry for the activities
of an interferometer. Clearly if our interferometers are operating
in a fuzzy (noncommutative) spacetime, rather than a sharp classical
spacetime, one should expect that at some point the accuracy
of the interferometer would be affected. This is indeed the
expectation of those working on noncommutative spacetimes,
but it has proven so far too hard to attempt to derive from
the noncommutative geometry the relevant physical effects.

\section{The test theories}

The study of the quantum-gravity problem has provided motivation
for the study of a rather large number of new effects.
In light of the nature of the indications that are coming
from the theory side it appears necessary to structure the
phenomenological efforts according to a few alternative
strategies.

In some cases a given quantum picture of spacetime or
a given quantum-gravity scenario definitely predicts a specific
effect, {\it i.e.} we can identify in the formalism an effect
which, if not found, could falsify the theoretical picture.
Then there is scope for a sharply aimed phenomenological effort,
characterized only by
the few parameters that the specific theoretical
picture involves.
A situation of this type is maturing in the study of certain
noncommutative spacetimes, perhaps most notably the $\kappa$-Minkowski
spacetime, whose single length parameter (here denoted by $\lambda$)
could soon be constrained at a beyond-Planckian level.

In other cases, especially in ambitious attempts at a complete solution
of the quantum-gravity problem, such as string theory and loop quantum
gravity, one is dealing with an extremely complex and rich formalism
which appears to have the potentiality of giving rise to a plethora
of new effects, even though one is not able (at least at present)
to establish that a certain new effect is definitely present
in the relevant theoretical framework.
This type of frameworks provide motivation for
a general approach for the description
of deviations from standard theories. Since even the most
basic principles of present-day physics are not necessary ``safe''
in these new theories, one should be prepared for
modifications of the fundamental equations describing physics.
The most fundamental equations are the equations governing the
standard model, that is, the gauge theory of the electroweak
and strong interaction and the Einstein field equation,
and the phenomenology should explore all candidate modifications
of these equations.

Concerning the target sensitivity that this phenomenology
should set for itself, obviously the Planck energy scale can
provide some tentative guidance, but it appears that
the phenomenology should be prepared for the possibility that
another, possibly lower, scale might characterize at least
some quantum-gravity effects. For those new effects, as in the example
of some modified dispersion relations mentioned earlier, whose description
automatically brings about a to-be-determined energy scale,
it is of course of primary importance from a quantum-gravity perspective
to find ways to test the hypothesis that the relevant energy scale
is indeed the Planck scale.
But even a lower sensitivity may turn out to be sufficient
to uncover a quantum-gravity effect.

This large phenomenological effort, which as a whole should
consider a large number of possible effects (and parametrizations
of the effects) and a large range of possible magnitudes for
these effects, must of course be structured in terms of some reference
test theories. Reference to some widely-adopted test theories
of course facilitates the comparison of sensitivities achieved
by different experiments, and the test theories
can also provide a language that bridges the gap between
experiments and
(usually rather formal) theory work on quantum gravity.
But even in setting up such test theories a few alternative
strategies should be explored simultaneously.
A key issue is whether the new effects end up taking the form
of new terms in the old type of formalisms or require even a new
formalism. Usually, even when a new formalism is required for the
full new theory, one can approximately incorporate the new effects
in the old formalism. And on the basis of this experience
it is natural to set up some test theories that indeed describe the
new effects via some new terms in the old type of equations.
But the phenomenology should also keep in mind that there have
been cases in the history of physics where the description
of the new physics really demanded a new formalism.
For example, in order to introduce the new effects of special
relativity in quantum mechanics one was tempted to formulate
a ``relativistic quantum mechanics'' (introducing new structures in
the old formalism) but it turned out to be necessary to
invent relativistic quantum field theory. We now understand
that there is a limit
of relativistic quantum field theory that can be described approximately
in terms of a ``relativistic quantum mechanics'', but the
nature of the limiting procedure by which the old formalism emerges
can only be established {\it a posteriori}, when the full theory
has been developed and understood.
The study of the quantum-gravity problem has confronted us with
new features like the mentioned IR/UV mixing, which for example
should act as a warning against assuming {\it a priori} that a
naive low-energy limit of quantum gravity should give us back
our present theories. Another example is the deformed
law of composition of momenta in $\kappa$-Minkowski spacetime, which is incompatible with
the standard field-theoretic setup and can only be switched off
when all other $\kappa$-Minkowski
modifications of Lorentz symmetry are switched off.
Even at very low energies, if one keeps the $\kappa$-Minkowski
dispersion relation (which could find room in the standard field-theory
setup), but neglects the modification of the law of composition
of momenta (which is not compatible with the standard field-theory
setup), inconsistent results are obtained, including some
significant {\it violations} of Lorentz symmetry (whereas $\kappa$-Minkowski should only involve a {\it deformation} of Lorentz symmetry).

So alongside the test theories that describe the new effects still in the old language, we should also develop some test theories that are prepared for the need of a new formalism. In practice the second type of test theories will need to simply avoid using too much of our present formalism. For example, as stressed here later in this Section, some phenomenology of Planck-scale departures from Lorentz symmetry focuses
on pure kinematics only, in order to avoid the assumption that dynamics should be described within our present formalisms. An example of the opposite type is the SME framework, where one considers a plethora of new effects, but all still codified according our present rules, including power-counting-renormalizability.

In this Section we give a partial overview of test theories that can be used in quantum-gravity phenomenology. We describe in some detail, since it was not covered in detail in other recent quantum-gravity-phenomenology reviews~\cite{Amelino-Camelia00a,Sarkar02,JacobsonLiberatiMattingly04,Amelino-Camelia04}, a phenomenological scheme for the modification of the Maxwell and the Dirac equation. And we describe more briefly, since pedagogical descriptions are already available in Refs.~\cite{Amelino-Camelia00a,Sarkar02,JacobsonLiberatiMattingly04,Amelino-Camelia04}, some other test theories that are relevant for the search of anomalous dispersion and of quantum-spacetime fluctuations. In any case, since the present theories are well proven by all current experiments, it is reasonable to introduce tests theories through small modifications of the present theories. 

A further important aspect of test theories is that, from an experimenter's point of view, detailed test theories also allow 'bookkeeping' of the possible effects that can be bounded by different experiments: All, even very different, effects are related to the same set of parameters. Only by the use of test theories one is able to 'compare' different experiments.

\subsection{The generalized Maxwell equation}\label{Sec:GeneralizedMaxwell}

\subsubsection{The model}

A very general model for generalized Maxwell equations is based on the assumption that the homogeneous Maxwell equations $dF = 0$ are still true. This can be based on a definition
of the electromagnetic field strength and the charge based on the Aharonov--Bohm
effect \cite{LaemmerzahlMaciasMueller05}.
In our approach the principles to formulate
generalized inhomogenous Maxwell equations are:
\begin{itemize}\itemsep=-2pt
\item linear in the field strength,
\item first order in the differentiation, and
\item small deviation from the standard Maxwell equations.
\end{itemize}
Therefore, the generalized Maxwell equation which we are
going to discuss and to confront with experiments is
\begin{equation}
\lambda^{\mu\nu\rho\sigma} \partial_\nu F_{\rho\sigma}
+ \chi^{\mu\rho\sigma} F_{\rho\sigma} = 4 \pi j^\mu \, .
\end{equation}
The requirement of this equations to describe a small
deviation from the standard theory leads
to $\lambda^{\mu\nu\rho\sigma}
= \eta^{\mu[\rho} \eta^{\sigma]\nu}
+ \chi^{\mu\nu\rho\sigma}$
where $\eta^{\mu\nu}$ is
the Minkowski metric diag(+ -- -- --) and,
in that frame, all components of $\chi^{\mu\nu\rho\sigma}$
and $\chi^{\mu\rho\sigma}$ are small compared to unity.
Therefore, all effects are calculated to first order in
these quantities only.

In our approach, the constitutive tensor $\lambda^{\mu\nu\rho\sigma}$
and, thus, the tensor $\chi^{\mu\nu\rho\sigma}$ are assumed to possess
the symmetry $\lambda^{\mu\nu\rho\sigma} = \lambda^{\mu\nu[\rho\sigma]}$
only (in the SME the constitutive tensor possesses the symmetry of the
Riemann tensor \cite{KosteleckyMewes02}).
The decomposition of this constitutive tensor and of $\chi^{\mu\rho\sigma}$
reads (see Ref.~\cite{LaemmerzahlMaciasMueller05} for the definition of the
various irreducible parts)
\begin{eqnarray}
\chi^{\alpha\beta\mu\nu} & = & {}^{(1)}W^{\alpha\beta\mu\nu} + \frac{1}{2} \epsilon^{\mu\nu}{}_{\rho\sigma} \eta^{\rho[\alpha} \Psi^{\beta]\sigma} + \frac{1}{12} X \epsilon^{\alpha\beta\mu\nu} - \eta^{\mu[\alpha} \Phi^{\beta]\nu} + \eta^{\nu[\alpha} \Phi^{\beta]\mu} \nonumber\\
& & + \eta^{\mu[\alpha} W_{\rm a}^{\beta]\nu} - \eta^{\nu[\alpha} W_{\rm a}^{\beta]\mu} - \frac{1}{6} W \eta^{\alpha [\mu} \eta^{|\beta|\nu]} + {}^{(1)}Z^{\alpha\beta\mu\nu} \nonumber\\
& &  + \frac{1}{2} \epsilon^{\mu\nu}{}_{\rho\sigma} \eta^{\rho(\alpha} \Upsilon^{\beta)\sigma} + \frac{1}{3} \left(2 \eta^{\mu(\alpha} \Delta^{\beta)\nu} - 2 \eta^{\nu(\alpha} \Delta^{\beta)\mu} - \eta^{\alpha\beta} \Delta^{\mu\nu}\right) \nonumber\\
& & + \frac{1}{4} \eta^{\alpha\beta} Z^{\mu\nu} + \frac{1}{2} \left(\eta^{\mu(\alpha} \Xi^{\beta)\nu} - \eta^{\nu(\alpha} \Xi^{\beta)\mu}\right)  \label{ConstGeneral} \\
\chi^{\mu\rho\sigma} & = & {}^{(1)}\chi^{\mu\rho\sigma} + \epsilon^{\mu\nu\rho\sigma} a_\nu + t^{[\rho} \eta^{\sigma]\mu} \, .
\end{eqnarray}

A 3+1 decomposition of the generalized Maxwell equations gives ($i, j = 1, 2, 3$)
\begin{eqnarray}
4 \pi \rho & = & \mbox{\boldmath$\nabla$} \cdot \mbox{\boldmath$E$} + \chi^{000i} \dot E_i + \chi^{00ij} \dot B_{ij} + \chi^{0i\rho\sigma} \partial_i F_{\rho\sigma} + \chi^{0\rho\sigma} F_{\rho\sigma}\label{Inhom1} \\
4 \pi j^i & = & \dot E_i - (\mbox{\boldmath$\nabla$} \times \mbox{\boldmath$B$})^i + \chi^{i00j} \dot E_j + \chi^{i0jk} \dot B_{jk} + \chi^{ij\rho\sigma} \partial_j F_{\rho\sigma} + \chi^{i\rho\sigma} F_{\rho\sigma} \, , \;\;\label{Inhom2}
\end{eqnarray}
where $E_i = F_{0i}$ and $B_i = \frac{1}{2} \epsilon_{ijk} F_{jk}$. Using the homogeneous Maxwell equations, the time derivative of $\mbox{\boldmath$B$}$ can be replaced by a spatial derivative
of the electric field. The appearance of the term $\chi^{000i}$ makes both equations to dynamical equations for the electric field rendering them to be an overdetermined system. Therefore we have to require the vanishing of the coefficient $\chi^{000i}$. Since this should be true for any chosen frame of reference, we have to require $\chi^{(\mu\nu\rho)\sigma} = 0$, what is identical to ${}^{(1)}Z_{(\alpha\beta\mu)\nu} = 0$, $\Xi_{\mu\nu} = 0$, and $\Delta_{\mu\nu} = - \frac{3}{4} Z_{\mu\nu}$.

Due to the vanishing of these irreducible parts we get $\partial_\mu(\lambda^{\mu\nu\rho\sigma} \partial_\nu F_{\rho\sigma}) = 0$ so that only $\chi^{\mu\rho\sigma}$ may lead to charge non--conservation:
\begin{equation}
4 \pi \partial_\alpha j^\alpha = \chi^{\alpha\mu\nu} \partial_\alpha F_{\mu\nu} = {}^{(1)}\chi^{\alpha\mu\nu} \partial_\alpha F_{\mu\nu} + \eta^{\alpha\mu} t^\nu \partial_\alpha F_{\mu\nu} \, .
\end{equation}

The validity of the homogenous Maxwell equations allows to require the vanishing of $\lambda^{\mu[\nu\rho\sigma]}$ without any physical consequences. This leads to $X = 0$, $Z^{\mu\nu} = W_{\rm a}^{\mu\nu}$, and $\Upsilon^{\mu\nu} = \frac{1}{2} \Psi^{\mu\nu}$.

\subsubsection{Radiation effects}

In order to discuss the effects due to the anomalous terms in radiation phenomena, we derive the wave equation
\begin{equation}
0 = \ddot E_i - \Delta E_i + (\mbox{\boldmath$\nabla$} (\mbox{\boldmath$\nabla$} \cdot \mbox{\boldmath$E$}))^i + \chi^{i\mu\nu j} \partial_\mu \partial_\nu E_j + 2 \chi^{i0j} \dot E_i + \chi^{ikj} \partial_k E_j \label{ModWaveEqn}
\end{equation}
and make a plane wave ansatz $\mbox{\boldmath$E$} = {\mbox{\boldmath$E$}}^0 e^{- i (\mbox{\boldmath$\scriptstyle k$} \cdot \mbox{\boldmath$\scriptstyle x$} - \omega t)}$ which results in equations for the amplidude and its derivative:
\begin{eqnarray}
0 & = & \left((\omega^2 - {\mbox{\boldmath$k$}}^2)\delta_{ij} + k_i k_j + 2 \chi^{i \mu j \nu} k_\mu k_\nu\right) E_j^0 \, . \label{0thorder} \\
0 & = & - 2 \omega \dot E^0_i - 2 k \cdot \nabla E^0_i + k_j \partial_i E^0_j + k_i \partial_j E^0_j \nonumber\\
& & + 2 \chi^{i\mu\nu j} \left(k_\mu \partial_\nu E^0_j + k_\nu \partial_\mu E^0_j\right) + 2 \omega \chi^{i0j} E^0_j - k_k \chi^{ikj} E^0_j \label{approx1} \\
0 & = & \ddot E^0_i - \Delta E^0_i + \partial_i \partial_j E^0_j + 2 \chi^{i\mu\nu j} \partial_\mu \partial_\nu E^0_j \, .
\end{eqnarray}
The first equation gives the dispersion relation
\begin{equation}
\omega = \left(1 + \rho(n) \pm \sqrt{\sigma^2(n) - \rho^2(n)}\right) |\mbox{\boldmath$k$}|
\end{equation}
with $\rho(n) = - \frac{1}{2} \Phi^{\mu\nu} n_\mu n_\nu$ and $\sigma^2(n) = \frac{1}{2} \eta_{\alpha\gamma} \eta_{\beta\delta} \chi^{\alpha\mu\nu\beta} \chi^{\delta\rho\sigma\gamma} n_\mu n_\nu n_\rho n_\sigma$, where $n_\mu = k_\mu/\omega = (1, \mbox{\boldmath$k$}/|\mbox{\boldmath$k$}|)$. This generalizes results in Ref.\ \cite{Ni74,KosteleckyMewes02}.

The $\pm$ sign in front of the square root indicates a hypothetical birefringence. If no birefringence is observed, then ${}^{(1)}W^{\mu\nu\rho\sigma} = 0$ and $\Psi^{\mu\nu} = 0$. For the interpretation of this result, see \cite{LaemmerzahlHehl04}. It has been shown \cite{KosteleckyMewes02} that from astrophysical observations birefringence can be excluded at the level of $10^{-32}$. The remaining $\rho(n)$ leads to an anisotropic speed of light which has been excluded in laboratory experiments \cite{Muelleretal03c} at the $10^{-15}$ level. Vanishing anisotropy implies $\Phi^{\mu\nu} = 0$. Furthermore, from (\ref{approx1}) a propagation equation for the amplitude can be derived
\begin{equation}
v^\mu \partial_\mu E_i^0 = - \left(2 \omega \chi^{i0j} - k_k \chi^{ikj}\right) E^0_j \, ,
\end{equation}
where $v^\mu$ is the group velocity of the light ray. This is directly related to the charge non--conservation parameter $\chi^{\mu\rho\sigma}$. Since no precession of the polarization has been inferred from astrophysical observation \cite{CarrollFieldJackiw90} the tensor $\chi^{\mu\rho\sigma}$ vanishes at the order $10^{-42}\;\hbox{GeV}$.

From all these requirements, the remaining generalized Maxwell equations are (we let aside the factor $W$ since this can be absorbed into a redefinition of the electric charge and current)
\begin{equation}
4 \pi j^\alpha = \partial_\nu F^{\mu\nu} - \frac{1}{2} \eta^{\alpha\mu} Z^{\nu\beta} \partial_\beta F_{\mu\nu} + \frac{3}{2} \eta^{\beta\mu} Z^{\nu\alpha} \partial_\beta F_{\mu\nu} + \frac{1}{2} \eta^{\alpha\beta} Z^{\mu\nu} \partial_\beta F_{\mu\nu}\, . \label{GME3}
\end{equation}
In the SME approach the $Z^{\mu\nu}$ are absent. Therefore, in our approach it is not possible to establish Lorentz invariance by radiation experiments only. Further experiments are needed.

\subsubsection{Electromagnetostatics}

The 3+1 decomposition of the above equations gives
\begin{align}
\frac{\rho}{\epsilon_0} & = - \mbox{\boldmath$\nabla$} \cdot \mbox{\boldmath$E$} - \hat{\mbox{\boldmath$\zeta$}} \cdot (\mbox{\boldmath$\nabla$} \times \mbox{\boldmath$E$}) - c \mbox{\boldmath$\zeta$} \cdot (\mbox{\boldmath$\nabla$} \times \mbox{\boldmath$B$}) \label{GenMax3p11} \\
\mu_0 \mbox{\boldmath$j$} & = \frac{1}{c^2} \dot{\mbox{\boldmath$E$}} - \frac{1}{c^2} \hat{\mbox{\boldmath$\zeta$}} \times \dot{\mbox{\boldmath$E$}} - (\hat{\mbox{\boldmath$\zeta$}} \cdot \mbox{\boldmath$\nabla$}) \mbox{\boldmath$B$} + \mbox{\boldmath$\nabla$} \times \mbox{\boldmath$B$}
- \frac{1}{c} \mbox{\boldmath$\nabla$} (\mbox{\boldmath$\zeta$} \cdot \mbox{\boldmath$E$}) + \frac{1}{c} \mbox{\boldmath$\zeta$} (\mbox{\boldmath$\nabla$} \cdot \mbox{\boldmath$E$}) \, , \label{GenMax3p12}
\end{align}
where we used SI units and defined $\zeta^i := \frac{3}{2} Z^{0i}$ and $\hat\zeta_i := \frac{3}{4} \epsilon_{ijk} Z^{jk}$.

The generalized Maxwell equations for a point charge at the origin are given by (\ref{GenMax3p11},\ref{GenMax3p12}) with $\rho = q \delta(r)$ and $\mbox{\boldmath$j$} = 0$.
With $\mbox{\boldmath$E$} = \mbox{\boldmath$\nabla$} \phi$ and $\mbox{\boldmath$B$} = \mbox{\boldmath$\nabla$} \times \mbox{\boldmath$A$}$ and the gauge $\mbox{\boldmath$\nabla$} \cdot \mbox{\boldmath$A$} = 0$ we get to first order in the perturbations
\begin{equation}
\phi = \frac{1}{4 \pi \epsilon_0} \frac{q}{r} \, , \qquad \mbox{\boldmath$A$} = \frac{q \mbox{\boldmath$\zeta$}}{4 \pi \epsilon_0 c r} \, .  \label{PointChargeAnsatz}
\end{equation}
This gives a magnetic field
\begin{equation}
\mbox{\boldmath$B$} = \frac{q}{4 \pi \epsilon_0 c} \frac{\mbox{\boldmath$\zeta$} \times \mbox{\boldmath$r$}}{r^3} \, . \label{zetamagnfield}
\end{equation}
Therefore, our model includes the feature that a point charge also creates a magnetic field.

If the source of the Maxwell equations is a magnetic moment $\mbox{\boldmath$m$}$ localized at the origin, then the Maxwell equations are  (\ref{GenMax3p11},\ref{GenMax3p12}) with $\rho = 0$ and $\mbox{\boldmath$j$} = \mbox{\boldmath$m$} \times \mbox{\boldmath$\nabla$} \delta(r)$.
We assume again a static situation and get to first order
\begin{equation}
{\mbox{\boldmath$A$}} = \frac{\mu_0}{4 \pi} \frac{\mbox{\boldmath$m$} \times \mbox{\boldmath$r$}}{r^3} \, , \qquad \phi = \frac{\mu_0 c}{4 \pi} \frac{\left(\mbox{\boldmath$\zeta$} \times \mbox{\boldmath$m$}\right) \cdot \mbox{\boldmath$r$}}{r^3} \, .
\end{equation}
A magnetic moment also creates an electric field of an electrical dipole with dipole moment $\mbox{\boldmath$d$} = \mu_0 c \epsilon_0 \mbox{\boldmath$\zeta$} \times \mbox{\boldmath$m$}$.
This feature is ''dual'' to the previous case.

For the new Lorentz invariance violating parameters $\mbox{\boldmath$\zeta$}$ and $\hat{\mbox{\boldmath$\zeta$}}$ there seem to exist no experimental results. However, we expect strong bounds on the $\zeta^i$ components from measurements with SQUIDs and from atomic spectroscopy. With SQUIDs weak magnetic fields of down to $10^{-14}\;\hbox{T}$ can be measured. If we assume that a measurement with SQUIDs of a magnetic field from a point charge does not lead to any magnetic field larger than the SQUID sensitivity, then, from $|\lambda \zeta / (2 \pi \epsilon_0 c \rho)| \leq 10^{-14}\;\hbox{T}$ for a line charge density $\lambda = 0.01 \;\hbox{C/m}$ at a distance of 1 cm, we get the estimate $|\zeta| \leq 2.7 \cdot 10^{-17}$. We strongly encourage experimentalists to carry out such an experiment.

The $\zeta^i$ will also lead to a hyperfine splitting, additional to the usual one. We get for the interaction Hamiltonian of an electron in the magnetic field (\ref{zetamagnfield}) of the nucleus
\begin{eqnarray}
H_{\zeta} & = & {\mbox{\boldmath$\mu$}}_{\rm el} \cdot \frac{q \mbox{\boldmath$\zeta$} \times \mbox{\boldmath$r$}}{4 \pi \epsilon_0 c r^3} \, .
\end{eqnarray}
If we choose the $z$--axis in direction of $\mbox{\boldmath$\zeta$}$, then the corresponding energy shift $\Delta E_{nlm} = \langle \psi_{nlm} \mid H_\zeta \mid \psi_{nlm}\rangle$ leads, e.g., to
\begin{equation}
\Delta E_{210} = - \frac{q \zeta_z \mu_x}{48 \pi \epsilon_0 c a^2} \, .
\end{equation}
With $\mu_z = e \hbar/m_e$ this yields $\Delta E_{210} = \zeta_z 1.8 \cdot 10^{-2}\;\hbox{eV}$. The state of the art of high precision measurements of energy levels is of the order $\Delta E/E \approx 10^{-15}$. Since the measured energy levels are still well described within the standard theory one gets for energies of about 10 eV at best an estimate $|\zeta_z| \leq 10^{-14}$ which, however, is not as good as a direct measurement discussed above might yield.

The parameter $\hat{\mbox{\boldmath$\zeta$}}$ gives rise to small deviations from the unperturbed quantities only and, thus, cannot be measured such precisely.

\subsection{The generalized Dirac equation}\label{Sec:GeneralizedDirac}

\subsubsection{The model}

The Dirac equation can be derived from the principles of (i) unique evolution, (ii) superposition principle, (iii) locality, (iv) conservation of probablility, and (iv) Lorentz invariance.
The requirements (i) -- (iii) leads to a linear system of partial differential equations of first order, requirement (iv) ensures the hyperbolicity of these equation. Using (i) to (iv) one gets
\begin{equation}
0 = i \gamma^\mu \partial_\mu \psi - M \psi
\end{equation}
where the matrices $\gamma^\mu$ in general do not fulfill a Clifford algebra, that is, there is no underlying Riemannian metric. As a consequence, the characteristic surface of this equation is given, as in the case of the Maxwell equation, by a forth order equation $0 = \det(\gamma^\mu k_\mu)$ and shows birefringence and anisotropy. The same phenomenon we encounter with the mass shell equation $0 = \det(\gamma^a p_a + M)$.

Also models with higher order derivatives as they appear in, e.g., in the effective equations for spin--$\frac{1}{2}$ particles within loop quantum gravity, have been
considered \cite{LaemmerzahlBorde01}.

\subsubsection{Propagation of Dirac particles}

There are two quasi--classical phenomena which one can derive from this generalized Dirac
equation \cite{Laemmerzahl98}: the equation of motion for the position in a nonrelativistic WKB aproximation
\begin{equation}
\dot a^i = - \left(\delta^{ij} + \alpha^{ij} + \bar\alpha^{ij}_k S^k\right) \partial_j U + \beta_{kl} \delta^{ij} \partial_j U^{kl}
\end{equation}
where $U$ is the Newtonian potential, $U^{ij}$ the Newtonian gravitational potential tensor \cite{Will93}, $S^i$ the spin and $\alpha^{ij}$, $\bar\alpha^{ij}_k$, and $\beta_{ij}$ are anomalous terms connected with the $X^{\mu\nu}$ and an anomalous coupling to $U^{ij}$,
see \cite{Laemmerzahl98}.
The corresponding equation for the precession of the spin is
\begin{equation}
\dot S^i = \epsilon^{ijk} \Omega_j S_k \qquad \hbox{with} \qquad \Omega_i = \tau^{kl}_i p_k p_l + \tau^j_i p_j + \tau_i\, ,
\end{equation}
where again the anomalous parameters $\tau^{ij}_k$, $\tau^i_j$, and $\tau_i$ are connected with the $X^{\mu\nu}$ and the anomalous coupling to $U^{ij}$.
In both cases the dynamics is influenced by anomalous parameters. The acceleration can be probed, e.g., by atom interferometry, the spin precession by the corresponding spin precession experiments performed, e.g., in high energy experiments.

\subsubsection{Spectroscopy}

For quantum particles also spectroscopy can give valuable information about the underlying dynamics of the fields. The anomalous parameters give rise to additional energy shifts and splitting of spectral lines. As an example, we mention the energy shifts occuring the Hughes--Drever like experiments (see Sec.\ref{Sec:ClockComparisonExp}) where a valence particle (a valence electron or a valence proton in the atomic's nucleus) leads to a splitting of the Zeeman
lines \cite{Laemmerzahl98} which, among others, yields the best estimates on an hypothetical anomalous inertial mass of the proton and on a hypothetical space-time
torsion \cite{Laemmerzahl97}.

Further effects like a modification of the spreading of a wave packet have been discussed in \cite{CamachoMacías04}.

\subsection{Generalized gravitational field}

In physics, geometry cannot be separated from the motion of physical objects. That something can be ''geometrized'' means that the dynamics of a certain object is independent from properties characterizing the object. In GR, the geometrization comes from the independence of the path of structureless particles from the mass and decomposition, the only characteristics of a structureless particle. The path is determined by its initial position and velocity only and, thus, given by an equation of the form $\ddot x^\mu + H^\mu(x, \dot x) = \alpha \dot x^\mu$. Then the motion is not related to the particular particles and, thus, can be assigned to an underlying geometry. The geometry underlying this geometrization of the path dynamics is called ''path structure''. This very general structure becomes more familiar when we assume that at each space-time point there is a frame so that the equation of motion for all particles reduces to $\ddot x^\mu = 0$ (Einstein's elevator). Then the equation of motion reads $\ddot x^\mu + H^\mu_{\rho\sigma}(x) \dot x^\rho \dot x^\sigma = \alpha \dot x^\mu$. The compatibility of this structure with the light cone structure then leads to a Weylian space--time which can be reduced to the ordinary Riemannian space-time by imposing the additional requiremet of the non--occurence of the second clock effect \cite{EhlersPiraniSchild72,Perlick87}.

The next step in establishing a theory for gravity then is to determine the equations which give the metric in terms of the matter content in the universe. In a particular case we get Einstein's field equations. A general parametrizations for a wide range of possible field equations for the metric is given by the PPN formalism, see e.g. \cite{Will93}. With the most important PPN parameters $\beta$ and $\gamma$, the metric is given by
\begin{eqnarray}
g_{00} & = & 1 - 2 U + 2 \beta U^2 \\
g_{0i} & = & V_i \\
g_{ij} & = & - \left(1 + 2 \gamma U\right) \delta_{ij}
\end{eqnarray}
where $U$ is thr Newtonian potential and $V_i$ is a gravitational field connected with a matter current. In this representation the metric is very well adapted for a confrontion with experimental data, e.g., from light deflection, redshift, gravitational time delay, perihelion shift, and data from binary systems. Einstein's field equations are characterized by $\beta = \gamma = 1$. Recently, it has been proposed in a string theory inspired dilaton model that $\beta$ and $\gamma$ may differ from unity by $10^{-9}$ and $10^{-5}$, respectively. The latter one is very close to the recent measurement by the Cassini mission \cite{BertottiIessTortora03} which gave $|\gamma - 1| \leq 2 \cdot 10^{-5}$.

\subsection{Anomalous dispersion}

As mentioned various quantum--gravity scenarios lead us to considering
Planck-scale modified dispersion relations of the general type (\ref{displeadbis}).
The derivation of the modified dispersion relation
can take a rather different path in different approaches.
For example in $\kappa$-Minkowski spacetime, where one has the
support of the deformed (Hopf) algebra, the dispersion relation
is obtained rather simply,
through a Casimir of the spacetime-symmetry algebra,
just like the classical dispersion relation is obtained through
the mass Casimir of the Poincar\'e (Lie) algebra.
But in some scenarios the argument may be rather involved,
as shown by the case of Loop Quantum Gravity, in which
we are still unable to perform a full
rigorous derivation of the dispersion relation.
This would require a sort of ``Minkowski vacuum'' for Loop Quantum
Gravity, which is still unknown. One therefore relies on some states
which are not the vacuum, but may reproduce some of the characteristics
of the vacuum. For example, some authors~\cite{GambiniPullin99,AlfaroMoralesTecotlUrrutia00}
have performed the analysis
using the so-called Loop-Quantum-Gravity ``weave states''.

A key point for the
development of test theories based on this modified dispersion
relations concerns the possibility
of assuming
that a field-theoretic formulation be admissible.
Some authors~\cite{MyersPospelov03,Jacobsonetal03} have relied on the field-theoretic setup,
in spite of the fact that it is inevitably nonrenormalizable
(within the field-theoretic setup Eq.~(\ref{displeadbis}) corresponds
to nonrenormalizable dimension-5 operators).
Other authors, perhaps the majority, have been concerned not only
with the nonrenormalizability of the corresponding field theory,
but also with the possibility that the relevant quantum-gravity
scenarios might have to be based rather significantly on decoherence,
something that field theory by construction cannot accommodate.
These other authors have preferred avoiding to commit to a
formulation of dynamics and are basing their phenomenology
exclusively on kinematics.

The pure-kinematics test theories only rely on the form of the
dispersion relation and the form of the law of energy-momentum
conservation, which is usually assumed to be unmodified.

A formal description of a field-theory-based test theory
hosting the modified dispersion relations (\ref{displeadbis})
is discussed in Refs.~\cite{MyersPospelov03,AmelinoCamelia02}.
For example, for a pure abelian gauge theory
one introduces the extra term
 \begin{equation}
 \frac{1}{E_P} n^a F_{ad} \mbox{\boldmath$n$} \cdot \mbox{\boldmath$\nabla$}(n_b {\tilde{F}}^{bd})
~,
\label{robeq}
\end{equation}
where $n_a$ is a background four-vector that codified the breakup of
Lorentz symmetry.

\subsection{Space--time fuzziness and decoherence}

The quantum-gravity literature on spacetime fuzziness (or ``spacetime foam'')
originates from a rich collection
of arguments~\cite{Padmanabhan87,DoplicherFredenhagenRoberts94,Ahluwalia94,NgVanDam94,AmelinoCamelia94,AmelinoCamelia96,Garay95},
combining general relativity and quantum
mechanics, which suggest that in the Planck-scale regime there should
be some absolute limitations on the measurability of distances.
It turns out to be most convenient~\cite{Amelino-Camelia04}
to characterize operatively this spacetime fuzziness as an irreducible (fundamental)
Planck-scale contribution to the noise levels in the readout of interferometers.
Interferometer noise can in principle be reduced to zero
in classical physics.
Ordinary quantum properties of matter already introduce an
irreducible noise contribution.
Spacetime fuzziness would introduce another irreducible source
of noise, reflecting the fact that the distances involved in
the experiment would be inherently unsharp in a
foamy spacetime picture.

While, as mentioned in the preceding sections, nearly all approaches
to the quantum-gravity problem predict some limitations on the accuracy
of localization, and therefore predict some spacetime fuzziness,
one is usually unable to rigorously derive from first principles a detailed
description of the physical consequences of this fuzziness, such as
the mentioned interferometric noise.
A phenomenology is being developed nonetheless,
exploiting~\cite{Amelino-Camelia04,AmelinoCamelia99,NgvanDam00,Amelino-Camelia01}
the fact that the role of the Planck scale in the needed formulas
has only a limited number of options, as a result of the
constraints introduced by dimensional analysis.

The key input needed for this phenomenology turns out
to be the power spectrum $\rho(f)$
of the Planck-scale-induced strain noise~\cite{AmelinoCamelia99}.
Combining some intuition about the stochastic-like features
of spacetime fuzziness and the dependence on the Planck length
one can rather easily reach a model of $\rho(f)$. For example if the effects depend linearly on the Planck length $L_p$ ($\equiv 1/E_p$) and the underlying phenomena are of random-walk type
one is inevitably led to
\begin{equation}
\rho_h \sim  L_p f^{-2} \Lambda^{-2} \equiv \zeta L_p f^{-2} L^{-2}
~, \label{noiserw}
\end{equation}
The proportionality to the square of the inverse of the frequency is a direct result of the assumption of random-walk-type processes. The length scale $\Lambda$ is needed on the basis of the dimensional analysis of the equation, and is to be treated as a free parameter to be constrained experimentally. One may choose to make reference to a dimensionless parameter, $\zeta$, which may be used to express the ratio of $\Lambda$ with the length $L$ of the arms of an interferometer or of an optical resonator.

Other hypothesis~\cite{Amelino-Camelia04,AmelinoCamelia99,NgvanDam00,Amelino-Camelia01} about the
stochastic-like features of the underlying Planck-scale processes
lead to other forms of $f$ dependence (and $L_p$ dependence)
of the strain noise power spectrum.
In general one should find ($\alpha > -1$)
\begin{equation}
\rho_h \sim \zeta_{\alpha\beta}  L_p^{1+\alpha} f^{-2-\beta}
L^{-2-\alpha-\beta}
~.
\label{noiseother}
\end{equation}

There is therefore some interest in
attempting~\cite{Amelino-Camelia04,AmelinoCamelia99,NgvanDam00,Amelino-Camelia01,Schilleretal04}
to improve limits on the parameters $\zeta_{\alpha\beta}$.

The idea of spacetime fuzziness also motivates some research
work on Planck-scale-induced decoherence.
It is in fact rather plausible that spacetime fuzziness might
affect the time evolution of quantum-mechanical states in such
a way that, for example, a pure state might evolve into a mixed state.
A Planck-scale decoherence scenario which has been extensively
studied, for what concerns the
phenomenological aspects,
is the one of Ref.~\cite{Ellisetal96} (and references therein)
which is motivated by the ``Liouville Strings''
approach~\cite{EllisMavromatosNanopoulos92,GravanisMavromatos01}
and describes the decoherence effects within a density-matrix formalism.

\section{Controlled laboratory experiments}

The realization that, as a result of the large astrophysical propagation
distances and (in some cases) the large energies of the particles
involved, in a few astrophysics contexts the analysis of quantum-gravity effects can achieve Planck-scale sensitivity has generated a large interest over the last few years,
as documented in Refs.~\cite{Amelino-Camelia00a,Sarkar02,JacobsonLiberatiMattingly04,Amelino-Camelia04} and references therein. However, the realm of astrophysics is also affected by some
key limitations, which encourage us to complement this phenomenology
with controlled/laboratory experiments.
In astrophysics we are just limited to the role of ``observers''
rather than having the opportunity to devise, set up and control
a measurement procedure. In interpreting astrophysical observations one is faced with the problem that the source as well as the space in between (that might contain gasses, other electromagnetic fields etc.) are not under control of the experimenter, and this will, in the majority of cases,
lead only to model-dependent results. Moreover, as mentioned above, the study of the quantum-gravity problem can motivate the search of a variety of new effects, which we would like
to study one by one, whereas in astrophysics we must take what Nature offers,
and this is not going to cover all aspects of the phenomenology of interest.

In this Section we discuss a few significant plans for
a quantum-gravity-phenomenology with controlled/laboratory experiments.
The description of examples of astrophysics studies is postponed to the next section, and will be structured
more succinctly, since several recent reviews
({\it e.g.}, Refs.~\cite{Amelino-Camelia00a,Sarkar02,JacobsonLiberatiMattingly04,Amelino-Camelia04})
have already covered in detail the approach based on astrophysics.

\subsection{Universality of Free Fall}

One of the first physical laws stated and subject to tests is the UFF. It began with the free fall tests of Galileo and its tests using a tilted plane. The best tests today use torsion balances and confirm the UFF at the level of $5 \cdot 10^{-13}$ \cite{Baessleretal99}. The best free fall tests reach the level of $10^{-10}$ \cite{KurodaMio89,KurodaMio90,NiebauerMcHughFaller87}. Future tests to be performed in space on dedicated satellite missions, namely MICROSCOPE and STEP, should be able to tests the UFF to a level of $10^{-15}$ and $10^{-18}$ respectively \cite{Touboul01a,Lockerbieetal01}. Therefore, MICROSCOPE as well as STEP are able to cover predictions from some quantum gravity motivated predictions.

\subsection{Universality of the Gravitational Redshift}

In these tests the ticking rates of two clocks is compared while the clocks change their position in a gravitational field. As in tests of the UFF, all pairs of clocks have to be considered. Presently, the best tests has been reported in \cite{BauchWeyers02} where two atomic clocks, an atomic fountain clock and a hydrogen maser, are subject to the changing gravitational field of the Sun. The two clocks show the same redshift to an accuracy of $1.5 \cdot 10^{-5}$. Other tests compared atomic clocks and clocks based on microwave and optical resonators and gave in both cases the same redshift within $2 \cdot 10^{-2}$ \cite{TurneaureStein88,Braxmaieretal02}.

\subsection{Lorentz invariance}

\subsubsection{Accurate, clean, and comprehensive tests}

Test of the principle of (local) Lorentz invariance (LI) are an extremely rich field for the search of fundamental physics. One reason is the universal validity of LI as a framework for all theories of nature. A violation of LI would thus manifest itself in virtually all branches of physics, and this allows to exploit the highest precision measurements as tests of LI. In test theories, the possible violations of LI are encoded into a set of parameters. The effects of these parameters in various experiments can be calculated (at least in principle), so that results on the parameters can thus be obtained from experiments. These theories have been advanced to a level where the outcome of all these experiments can be compared. This also answers the question whether two experiments test different aspects of relativity or whether they are different methods for observing the same effect.

Ideally, the experimental verification of LI should be {\em accurate, clean,} and {\em comprehensive.} Accurate refers to the ability to detect very small violations and clean refers to how well the hypothetical effect sought in experiments is understood and referred to a set of parameters in the test theory. This is a requirement to both the experiment and the underlying theory. It ensures that a violation of relativity does not accidentally cancel. Comprehensiveness denotes the degree of completeness to which the theoretically conceivable violations of LI are excluded.

This summary wants to present the most important experimental techniques that have been used for bounding violations of LI. The variety of methods makes this an interesting and challenging subject for experimental physicists. As a general observation, tight limits on spin-dependent parameters can be obtained by comparing similar particles having different polarizations. The spin-independent effects have to be measured by comparing to independent standards, e.g., a macroscopic body as a length standard. Another common feature of most experiments is that they seek for a Lorentz-violating dependence of an effect as a function of the orientation of the apparatus or the velocity of the laboratory frame in space. For a laboratory located on Earth, the relevant rotations and velocities can be given by Earth's orbit as well as rotation.

\subsubsection{Applications and every day tests}

LI has found many applications in fundamental and applied research, in technology and, thus, even in every day life. One striking example on how every day these may be are the accelerated electrons in cathode ray tubes used in color television sets: Their mass is increased by about 5\% relative to their rest mass. For a less mundane example, the accurate quantum field theoretical prediction of the electron's anomalous magnetic moment $g=2.002\ldots$ to twelve significant digits \cite{Kinoshita96} (maybe the most accurate prediction of a 'complicated' number in any science) also is a confirmation of LI.

In the global positioning system, a receiver determines its position by comparing timing signals received from accurate reference clocks located in satellites. The positions of the satellites are known, and the difference in the timing signals gives the relative distances that allows the receiver to determine its spatial coordinates and the local time. However, since the reference clocks are moving with respect to the receiver, their rates are subject to time dilation (and also to the redshift due to Earth's gravity). Not taking into account these effects would lead to a position error of many kilometers \cite{Bahder03}.

There are also technologies whose very operating principle is a relativistic effect. Certain microwave oscillators ('gyrotrons') \cite{MeinkeGundlach86} and the free electron laser \cite{OSheaFreund01} use the relativistic Doppler effect to generate short-wavelength radiation. In a free electron laser, a beam of electrons is directed along an arrangement of periodically poled magnets, where the period $p$ is of the order of a centimeter. The moving electrons have a kinetic energy $E\gg m_e c^2$, and thus see a Lorentz-contracted period $p m_e c^2/E$. The corresponding oscillating  magnetic field in the electron's rest frame thus has a frequency $\nu_e=cE/(p m_e c^2)$ and forces the electrons to an oscillatory motion that leads to the emission of electromagnetic waves with a frequency $\nu_e$. The Lorentz transformations back to the laboratory frame concentrate these waves into a tight cone in the forward direction of the electron beam, with a blue-shifted frequency $\nu_{\rm lab} = \nu_e E/(m_e c^2)=cE^2/(p m_e^2 c^4)$. Thus, highly concentrated radiation and a small wavelength $\lambda =p(m_e c^2/E)^2$ are achieved. The two Lorentz transformations give the quadratic dependence on $1/E$.

Particle accelerators are maybe the most prominent example of a technology where relativistic effects are built in. The velocity-dependence of the relativistic mass, for example, is taken into account in the construction of the machines, and, most importantly, the understanding of the high-energy particle reactions provides maybe the richest field for the application of relativistic (quantum field) theories. These have not only given rise to an enormous output of fundamental physics discoveries, but the apparent correctness of relativistic quantum field theory is also a confirmation of relativity itself.

While the success of relativistic physics in fundamental and applied research gives us confidence into the theory, the 'every-day tests' provide no high-precision confirmation: On the one hand, machines like particle accelerators, microwave oscillators etc. are constructed in such a way that small tolerances do not lead to malfunction, and this also means that small errors in the underlying theory may not affect functionality. Also, the 'every day tests' are usually not clean: In accelerator experiments, for example, an incomplete understanding of the hadronic processes might cover a tiny violation of relativity. For these reasons (and for achieving comprehensiveness), LI has been tested in dedicated experiments since its inception.

\subsubsection{Macroscopic matter effects}\label{Sec:MacroscopicMatterEffects}

Lorentz violation affects the properties of macroscopic bodies through a modification of its microscopic constituents. These modifications, in particular the change of the geometry, have been studied for the interpretation of experiments. As an important result, within the SME matter effects do not cancel the sensitivity in interferometer or cavity tests of Lorentz invariance. For a model which may lead to a cancellation see \cite{LaemmerzahlHaugan01}. Instead, they enhance the sensitivity for Lorentz violation in electrodynamics, but only slightly for cavity materials presently in use. Moreover, the theory concerning the Dirac sector \cite{Muelleretal03d} allowed to derive the first experimental limits on some of the electron coefficients $c_{\mu\nu}$, at a level of $10^{-14}$. The theories summarized here constitute a complete description of all SME effects that influence experiment using vacuum-filled cavities. These experiments are thus particularly clean tests of Lorentz invariance.

\paragraph{Photon sector}

Within the electrodynamic sector of the SME, Lorentz violation leads to a modified velocity of light $c=c_0+\delta c$ and also to a modified Coulomb potential of a point charge $e$, $\Phi(\vec x)=e^2/(4\pi |\vec x|)+ V$, where \cite{KosteleckyMewes02}
\begin{equation}\label{Coulomb}
V=\frac{e^2}{8\pi} \frac{\mbox{\boldmath$x$} \cdot \kappa_{DE} \cdot \mbox{\boldmath$x$}}{|\mbox{\boldmath$x$}|^3}\,.
\end{equation}
The influence of this on solids can be treated for ionic crystals \cite{Muelleretal03}, which in the simplest case (e.g., NaCl) consists of a lattice of ions with opposite charges. The lattice is formed by the balance between attractive Coulomb forces and a quantum mechanical repulsion due to the overlap of the ionic orbitals. Perturbative calculations show that the change in the repulsive potential due to Lorentz violation is negligible. For estimating the influence of the modification of the Coulomb potential $V$, the force generated by the modified Coulomb potential is summed over all ions and equated to the elastic force associated with a distortion of the lattice. This leads to the relative change of the length
\begin{eqnarray} \label{lengthchange}
\frac{\delta L_z}{L} & = & A
\left[(2\sigma-3\tau_\|) (\kappa_{DE})_{\|}-3\tau_{\perp}(\kappa_{DE})_\perp \right]
\end{eqnarray}
with $\sigma$ and $\tau$ being constants obtained from summing the Coulomb potential in analogy to the Madelung constants. $(\kappa_{DE})_\|=(\kappa_{DE})_{zz}$,
$(\kappa_{DE})_\perp=(\kappa_{DE})_{xx}+(\kappa_{DE})_{yy}$ and
\begin{equation}\label{afaktor}
A=-\frac 12 \frac{e^2 v_1 v_2}{8\pi E_{\rm Y}} \frac{N_m N_A \rho}{M a}
\, .
\end{equation}
$v_1$ and $v_2$ are the number of valence charges for the atoms, and $E_Y$ is the elastic modulus. The contributions of the length change and the change in the speed of light give the total frequency change of a cavity filled with vacuum due to Lorentz violation in electrodynamics
\begin{equation}\label{deltanuaddl}
\frac{\delta \nu_{\rm cav}}{\nu_{\rm cav}} = -a_\| \hat N \kappa_{DE}\hat N -
\left(\frac12+a_\perp \right) \left[\hat E^*\kappa_{DE}\hat E + (\hat N\times \hat
E^*)\kappa_{DE}(\hat N\times \hat E)  \right] \, .
\end{equation}
Here, $a_\|=A(2\sigma-3\tau_\|)$ and $a_\perp=-3A\tau_{\perp}$. This has been simplified by noting that astrophysical tests lead to $(\kappa_{HB})=-(\kappa_{DE})$. For practical materials sapphire and quartz the length change effect is negligible (see Table \ref{Avalues}). For future experiments using resonators made of other materials, however, the influence might be stronger and {\em enhances} the sensitivity.

\begin{table}
\centering
\caption{\label{Avalues} Length change coefficients.}
\begin{tabular}{crrcrr}
\hline
Material & $a_\|$ & $a_\perp$ & Material & $a_\|$ & $a_\perp$
\\   \hline NaCl & -0.28 & 0.10 & LiF & -1.06 & 0.37
\\ sapphire & -0.03 & 0.01 & quartz  & -0.11 & 0.04 \\ \hline
\end{tabular}
\end{table}

The model can be extended to include the additional non-Lagrangian coefficients of the most general Maxwell equations that are linear and first order in the
derivatives \cite{LaemmerzahlMaciasMueller05,MuellerLaemmerzahl04}. It is found that the non-SME terms do not additionally modify the geometry of crystals.

\paragraph{Fermionic sector}

Here, the starting point is the nonrelativistic hamiltonian $h=h_0+\delta h$ of a free electron in the SME, as described in \cite{KosteleckyLane99}. Most of the terms contained therein (Tab. \ref{clockparameters}) have zero expectation value within the rest frame of a solid. The only term that doesn't drop out for non-spin-polarized materials (spin-polarized materials can also be treated\cite{Muelleretal03d}) is the modification of the kinetic energy of the electron, $\delta h=E'_{jk}p_jp_k$. Since the electrons inside crystals have a nonzero expectation value $\left<p_ip_j\right>$, which is a function of the geometry of the lattice, Lorentz violation will cause a geometry change ('strain') of the crystal.

Strain is conventionally expressed by the strain tensor $e_{ij}$. For $i=j$, it represents the relative change of length in $x_i$-direction, and for $i \neq j$, it represents the change of the right angle between lines originally pointing in $x_i$ and $x_j$ direction. A general linear relationship between the Lorentz violating quantity $E'_{jk}$ and strain is given by
\begin{equation}\label{edctensor}
e_{dc} = \mathcal B_{dcpj} E'_{pj}.
\end{equation}
with a 'sensitivity tensor' $\mathcal B_{dcpj}$ that has to be determined from a model of the crystal. $\mathcal B_{dcpj}$ can be taken as symmetric in the first and last index pair; symmetry under exchange of these pairs will hold only for some simple crystal geometries, like cubic. Thus, the tensor has at most 36 independent elements. To calculate the sensitivity tensor, the electronic states are described by Bloch wave functions to determine $\left<p_ip_j\right>$; the corresponding strain is calculated using elasticity theory. As a result, the sensitivity tensor $\mathcal B_{dcjp} =  \mu_{dcmp} \kappa_{mj}+\mu_{dcmj} \kappa_{mp}$,
where
\begin{equation}\label{kappadef}
\kappa_{mj}= \frac{N_{e,u}  \hbar^2}{m |\mbox{det}(l_{ij})|} k_{ml} k_{jk} {\overline\xi}_{lk} \,,
\end{equation}
can be calculated. $N_{e,u}$ is the number of valence electrons per unit cell, $|$det$(l_{ij})|$ is the volume of the unit cell expressed by the determinant of the matrix of the primitive direct lattice vectors, $k_{ml}$ is the matrix containing the primitive reciprocal lattice vectors, and $\mu_{dcmp}$ the elastic compliance constants. The symmetric $3\times 3$ matrix ${\overline\xi}_{lk}$ is given by the Fourier coefficients of the Bloch wave functions. Its six parameters are unknown at this stage and can, e.g., be determined from a simple model that leads to ${\overline\xi}_{lk} \sim \delta_{lk}$. To eliminate these unknowns, an alternative method is used to calculate the strain for the simple case of isotropic Lorentz violation $E'_{jk}\sim \delta_{jk}$, and the result is compared to Eq. (\ref{edctensor}) \cite{Muelleretal03d}. This yields six equations from which $\kappa_{ab}$ (that depends solely on material properties) can be determined and re-inserted into Eq. (\ref{edctensor}):
\begin{equation}\label{sensnew}
\mathcal B_{dcjp} =
\mu_{dcmp}\lambda_{aamj}+\mu_{dcmj}\lambda_{aamp}\,.
\end{equation}
Note that the theory now needs no assumptions that go beyond the use of Bloch states. Thus, it is very generally applicable and accurate.

For convenience, we arrange the independent elements of $e_{ab}$ into a six-vectors $\mathbf e= (e_{xx}, e_{yy}, e_{zz}, e_{yz}, e_{zx}, e_{xy})$ and express Eq. (\ref{edctensor}) as $\mathbf e=\mathcal B \cdot \mathbf E'$, where $\mathcal B$ is a $6\times 6$ 'sensitivity matrix'. Tab. \ref{elektronenterme} gives values for the materials presently used for cavities. Gold is included as an example for a material with exceptional sensitivity.

The influence of the electron terms $c_{\mu\nu}$ on hydrogen molecules have also been calculated \cite{Muelleretal04}. Here, an explicit wave-function can be obtained from first principles, and Lorentz-violating changes in the frequencies of electronic and (ro-) vibrational transitions, as well as the bond length, have been obtained. This allows new tests of Lorentz symmetry that use molecules. A change in the index of refracion of dielectric media is also connected to violation of LI \cite{Mueller04}. It has to be taken into account for certain cavity experiments.

\begin{table}
\centering
\caption{Sensitivity coefficients.\label{elektronenterme}}
\begin{tabular}{crrrrrrrr}\hline
Mat. & $\mathcal B_{11}$ & $\mathcal B_{12}$ & $\mathcal B_{13}$ &
$\mathcal B_{14}$ & $\mathcal B_{31}$ & $ \mathcal B_{33}$ &
$\mathcal B_{41}$ & $\mathcal B_{44}$ \\ \hline
Au & 24.13  & -11.06    &&&&&& 12.34 \\
  Al$_2$O$_3$ & 3.58 & -1.05 & -0.53 & 0.014 &  -0.57 & 3.14 & 0.004 &  5.08  \\ Nb & 6.80  & -2.40 &   &&&&& 17.9 \\ fused quartz & 2.64 & -0.32 &&&&&& 3.95\\
\hline
\end{tabular}
\end{table}

\subsubsection{Cavity experiments}\label{SMEcavexp}

Polarization experiments, see Sec.\ref{Sec:ObservationsInAstrophysics}, comparing light with light, cannot measure all 19 coefficients of the ordinary constitutive tensor $(k_F)_{\kappa \lambda \mu \nu}$, whereas cavity and interferometer experiments, comparing light with matter, can. Moreover, cavity experiments can measure coefficients for Lorentz violation in the electron's equation of motion that have not been measured by other methods. Cavity experiments have been developed out of the well known interferometer experiments originally performed by Michelson, Morley \cite{Michelson81,MichelsonMorley87} as well as Kennedy, Thorndike \cite{Kennedy26,KennedyThorndike32}, and others.

The basic principle is to measure the time of flight of light rays that transverse a path, which is usually defined by a pair of mirrors and a spacer. A change connected to the orientation or the velocity of the apparatus in space would indicate Lorentz violation. Interferometers indicate such a change by a shift in fringes formed by the interference of two beams. In cavities, however, interferometry between multiple beams is used. Each transverses the cavity a large number of times that is limited by losses such as imperfect mirrors. Today, about $10^5$ reflections are used, and $\sim 10^6$ seem technically possible. Thus, a large effective distance of propagation is reached in a compact apparatus, that can be much better shielded from temperature fluctuations and (seismic) vibrations, and is less sensitive to gravitational bending\footnote{In the large interferometers dedicated to the search for gravitational waves, shielding against seismic vibrations with frequencies above a couple of Hz is effectuated by pendulum-like mounting of the interferometer end mirrors. Thus, however, the dc position of the mirrors is much less stable than in the rigid cavities described here.}.

In cavity experiments, one measures the resonance frequencies
\begin{equation}
\omega=2\pi\frac{ mc}{2nL}
\end{equation}
($m$ is a constant mode number, $c$ is velocity of light parallel to the cavity axis, $n$ the index of refraction of the medium, and $L$ the cavity length) of a cavity, defined by the boundary conditions for standing waves. A Lorentz-violating shift in $c$ and/or $L$ and $n$ connected to a rotation or boost of the apparatus would lead to a measurable shift in $\omega$. The shift in $c$ is, of course, due to Lorentz violation in the photonic sector. The shift in $L$ has been determined above.

\begin{table}
\caption{\label{cavitytable} Overview of recent cavity experiments (WGR = whispering-gallery resonator). The notation $1-2 (4)$ in the column for $\kappa_{e-}$ indicates that the experiment provides 4 independent limits on parameters of $\tilde\kappa_{e-}$, the lowest individual being about 1, the highest about 2 parts in $10^{15}$ (absolute mean plus standard error). If no number in brackets is given, all matrix elements have been limited. The notation $\sim 5$ indicates that, although a detailed analysis within the SME was not performed, the experiment should limit at least one of the parameters at the indicated level.}
\begin{center}
\begin{tabular}{llrrr}\hline
Ref. & type & $\tilde\kappa_{e-}$ & $\tilde\kappa_{o+}$ & $c_{AB}$ \\
& & $\times 10^{15}$ & $\times 10^{11}$ & $\times 10^{15}$ \\ \hline
\cite{BrilletHall79} (1979) &  rotat. quartz cavity/CH$_4$ & $\sim 6$ & $\sim 10$ & \\
\cite{HilsHall90} (1980) & quartz cavity/CH$_4$ & & & \\
\cite{Braxmaieretal02} (2000) & sapphire cavity/iodine & & & \\
\cite{Lipaetal03} (2003) & Nb microwave cavities & 140-430 (4)& 200 (1) & \\
\cite{Muelleretal03c} (2003) & sapphire cavities & 4-18 (4)& 3-28 & \\
\cite{Muelleretal03d} (2003) & reanalysis of \cite{BrilletHall79,Muelleretal03c} & 4-18 (4) & 3-28 & 2-100 (3) \\
\cite{Wolfetal04a} (2004) & sapphire WGR/Cs clock & 2-8 (4) & 3-5 & \\ \hline
\end{tabular}
\end{center}
\end{table}

\begin{table}
\centering
\caption{\label{cavityresults} Limits from recent cavity experiments.}
\begin{tabular}{lrr}\hline
Parameter & \cite{Muelleretal03c,Muelleretal03d} & \cite{Muelleretal03c,Wolfetal04a,Mueller04}\\
\hline
$(\tilde\kappa_{e-})^{XY}/10^{-15}$ & $1.7\pm2.6$ & $-1.7\pm1.6$ \\
$(\tilde\kappa_{e-})^{XZ}/10^{-15}$ & $-6.3\pm 12.4$ & $-4.0\pm3.3$ \\
$(\tilde\kappa_{e-})^{YZ}/10^{-15}$ & $3.6\pm 9.0$ & $0.5\pm 2.52$ \\
$((\tilde\kappa_{e-})^{XX}-(\tilde\kappa_{e-})^{YY})/10^{-15}$ & $8.9\pm
4.9$ & $2.8\pm3.3$ \\
$(\tilde\kappa_{o+})^{XY}/10^{-11}$ & $14\pm14$ & $-1.8\pm 1.5$ \\
$(\tilde\kappa_{o+})^{XZ}/10^{-11}$ & $-1.2\pm 2.6$ & $-1.4\pm 2.3$ \\
$(\tilde\kappa_{o+})^{YZ}/10^{-11}$ & $0.1\pm 2.7$ & $2.7\pm 2.2$ \\
$c^e_{(YZ)}/10^{-15}$ & & $0.21\pm0.46$ \\ $c^e_{(XZ)}/10^{-15}$ & &
$-0.16\pm0.63$\\
$c^e_{(XY)}/10^{-15}$ & $\lesssim 1$ & $0.76\pm0.35$\\
$(c^e_{XX}-c^e_{YY})/10^{-15}$ & $\lesssim 8$ & $1.2\pm0.64$\\
$|c^e_{XX}+c^e_{YY}-2c^e_{ZZ}-0.25(\tilde\kappa_{e-})^{ZZ}|/10^{-12}$ &
$\lesssim 1$ & $\lesssim 1$ \\ \hline
\end{tabular}
\end{table}

The first experiment to make full use of the potential that reference cavities offer for precision experiments was performed by Brillet and Hall \cite{BrilletHall79} (see Tab. \ref{cavitytable}). A laser was stabilized to a resonance of a fused quartz cavity (actually, ultra-low expansion (ULE) glass ceramics). Both the cavity as well as the laser were rotating on a platform. Their frequency was compared to a stationary methane (CH$_4$) frequency standard. The data showed frequency variations of a few parts in $10^{13}$ that where ascribed to a slight tilt of the rotation axis with respect to Earth's gravity. After a transformation of the signal into the sidereal frame, however, this effects becomes oscillatory and can thus be separated from any signals of cosmic origin. The experiment was performed before the developement of an elaborate theoretical framework. A detailed re-analysis of the published result within the SME is, unfortunately, not possible, since only a single signal component is given. For deriving detailed SME results, it would be desirable to have at least 10 independent components.

In a smilar setup, a non-rotating quartz cavity (ULE) was compared against a ${\hbox{CH}}_4$ standard by Hils and Hall \cite{HilsHall90}. The signal period was given by Earth's rotation. On that time-scale, however, the ULE cavity showed a significant drift that limited the accuracy of that experiment. The experiment was interpreted in the RMS test theory (Sec.\ref{Sec:KinematicalTestTheories}) and provided the best limits of $6.6\times 10^{-5}$ on the velocity-dependence parameter at the time. As above, a detailed analysis within the experiment from the published data is not feasible. This experiment was improved by a factor of about three by Braxmaier {\em et al.} \cite{Braxmaieretal02}, who used a cavity made from crystalline sapphire, operated at the temperature of liquid helium (4\,K). The crystalline material showed a remarkable absence of creep, i.e., no systematic long term drifts of the cavity length. Thus, it was possible to search for a signal with 1\,year period, given by Earth's orbit.

The first experiment to determine simultaneous limits on the photonic SME coefficients was published in 2003 \cite{Lipaetal03}. Two superconducting microwave cavities made of Niobium (Nb) were used to give separate limits on four components of $\tilde\kappa_{e-}$ and one limit on $\tilde \kappa_{o+}$ (Tab. \ref{cavitytable}). Although the accuracy of this experiment was lower than that of older experiments, it determined simultaneous limits on several parameters, thus representing progress towards a comprehensive verification of LI.

The experiment of M\"uller {\em et al.} \cite{Muelleretal03c,Muelleretal03f}, improved both the accuracy as well as the coverage of parameter space. It determined four elements of $\kappa_{e-}$ to a level of about $10^{-15}$. The three elements of $\tilde\kappa_{o+}$ enter the signal suppressed by the velocity of Earth's orbit, $\beta\sim 10^{-4}$, and where thus limited to about $10^{-11}$ (Tab. \ref{cavityresults}). That one of these seven parameters was different from zero at 1.8$\sigma$ does {\em not} mean that Lorentz violation was detected. Rather, the probability of a mean value to be larger than the standard error is 23\% even if the true value is zero, so  one or two such instances are to be expected if the error bars are realistic. The experiment compared two cryogenic sapphire resonators pointing in orthogonal directions over a period of 399 days. This made it possible to independently measure Fourier signal components separated by as little as 1/1\,year in frequency space, and thus allowed the experiment to give separate limits on the elements of $\tilde\kappa_{o+}$.

The experiment of Wolf {\em et al.} used a slightly different setup, in which resonances of a cryogenically cooled microwave whispering gallery resonator (WGR) are used. The WGR is a cylinder made from crystalline saphirre, in which the microwave mode travels along the perimeter, guided by total internal reflection. Also this experiment was operated for a sufficiently long time to state separate limits on the elements of $\tilde \kappa_{o+}$. In this experiment, the results for the three parameters are significant at about 2$\sigma$. As explained above, however, this is to be expected from standard statistics even if all parameters are zero, and thus not an indication of Lorentz violation.

The theory of the influence of fermionic LI violation in the cavity geometry \cite{Muelleretal03d} made it possible to state the first limits on three electron parameters contained in $c_{\mu\nu}$. Since this influence is material dependent, it can be separately measured by perfoming experiments with different cavity materials, here the experiments of Brillet and Hall \cite{BrilletHall79} and M\"uller {\em et al.} (Tab. \ref{cavityresults}). Theoretically, all components of $c_{\mu\nu}$ can be limited by this method, if sufficiently detailed experimental results are available. As explained above, this is unfortunately not the case for the older experiments.

For a complete determination of the photon and electron coefficients, the
experiment of M\"uller {\em et al.} has to be compared to an experiment that
gives as many signal components, but uses dissimilar cavities. The
experiment by Wolf {\em et al.} \cite{Wolfetal04a} could be used for this
purpose. However, in the whispering-gallery type cavity, 98\% of the
electromagnetic field energy are confined within a refractive material.
Thus, a shift in the index of refraction also connected to Lorentz violation
alters the Lorentz-violation signal. This has been taken into account in
\cite{Muelleretal04}, leading to more comprehensive and more accurate limits (Tab.
\ref{cavityresults}). The additional hypothetical shift in the index of
refraction of sapphire contributes to the higher accuracy.

\subsubsection{Doppler shift experiments}

To measure the relativistic Doppler shift, the frequency of a moving oscillator is measured as a function of the velocity with respect to the detector. Such measurements can be easily described within the RMS framework, where a parameter $\alpha$ describes the strength of the second order Doppler shift. $\alpha=-1/2$ is predicted by SR. In the SME and other dynamical test theories, the analysis is more complicated, since it must contain the nature of the moving clock. The most precise test yet gives $|\alpha_{MS}+1/2| \lesssim 1.8 \times 10^{-7}$ \cite{Saathoffetal03} by spectroscopy on lithium ions moving within a storage ring. The accuracy of this result is limited by the knowledge of the rest-frame transition frequency of lithium.

Other methods to measure the Doppler shift include $\gamma$-ray spectroscopy using a rotating M\"ossbauer absorber \cite{ChampeneyIsaakKhan65}. While the resolution of the frequency measurement is much higher, the velocity is limited to maybe 100 m/s due to centrifugal forces in the rotor, making these experiment less accurate.

In the SME and other dynamical test theories, the interpretation of such experiments is different: Here, the physics of the clock and the detector is described by dynamical equations, from which the received frequency can be calculated. A departure from the relativistic Doppler shift would arise from the extra terms in the dynamical equations. Thus, the analyis must contain the nature of the moving clock. Although this was already discussed \cite{Kostelecky05}, at the time of this writing no such analysis has been performed yet.

\subsubsection{Clock-comparison experiments}\label{Sec:ClockComparisonExp}

The basic idea is to compare the oscillation frequencies of two physical systems (``clocks") and
look for any change with the orientation or velocity of the clocks in space. The highest precision
limits are achieved by comparing the frequencies of dissimilar clocks as they rotate with the
Earth.

If the influence of the Lorentz-violating parameters on the frequencies of the clock can be calculated (by use of perturbation theory), definite bounds can be obtained by a comparison to the outcome of the experiment. However, the parameters of the SME are so manifold that a calculation of the individual influences is generally difficult, especially if nuclear energy levels are involved. In these cases, the problem has been simplified by use of simplified models for nuclear matter and assuming that all but a few parameters for Lorentz violation are zero \cite{KosteleckyLane99}. This gives insight into how many of the parameters can be bounded at which level of accuracy. Further assuming no cancellations between the parameters, limits on many parameters for Lorentz violation that lead to spin-dependent effects for the electron, the proton, and the neutron could be obtained (Tab. \ref{clockparameters}).

\begin{table}
\centering
\caption{\label{clockparameters} Combinations of Lagrangian parameters and the current order-of magnitude bounds (expressed as the decimal logarithm), obtained by using assumptions described in the text (from \cite{KosteleckyLane99}). $c_{(XY)}=\frac 12(c_{XY}+c_{YX})$.}
\begin{tabular}{lccc}\hline
parameter & proton & neutron & electron \\
 & GeV & GeV & GeV \\ \hline
$\tilde b_J=b_J-md_{J0}+\frac 12 \epsilon_{JKL}(g_{KL0}-H_{KL})$ & -27 & -30 & -27 \\
$\tilde c_{Q}=m(c_{XX}+c_{YY}-2c_{ZZ})$ & & & \\
$\tilde c_{Q,J}=2mc_{(JZ)}$ ($J=X,Y$) &  & -25 &  \\
$\tilde c_{-}=m(c_{XX}-c_{YY})$ & & -27 &  \\
$\tilde c_{XY}=2mc_{(XY)}$ & -27 & & \\
$\tilde d_J=2md_{(J0)}-\frac 12 (md_{J0}+\frac 12 \epsilon_{JKL}H_{KL})$ & -25 & -28 & -22 \\ $\tilde g_{D,J}=m\epsilon_{JKL}(g_{K0L}+\frac 12 g_{KL0})-b_J$ & -25 & -28 & -22 \\
$\tilde g_Q=m(g_{X0X}+g_{Y0Y}-2g_{Z0Z})$ & & & \\
$\tilde g_{Q,J}=m(g_{J0Z}+g_{Z0J})$ & & & \\ $\tilde g_{-}=m(g_{X0X}-g_{Y0Y})$ & & & \\ $\tilde g_{XY}=m(g_{X0Y}+g_{Y0X})$ & & & \\ \hline
\end{tabular}
\end{table}

In the tests originally performed by Hughes \cite{Hughesetal60} and Drever \cite{Drever61} at the beginning of the sixties, the clocks are based on transitions in atoms or ions between states characterized by the orientation of a electronic or nuclear spin (hyperfine or Zeeman transitions). If the rotation--invariance of the energy $E=mc^2$ of the nucleus was violated, the energy levels of the states involved would depend on the orientation of the clock in space. This would cause a modulation of the clock frequency $\nu_1$ with its orientation, that can be detected by comparing it to the frequency $\nu_2$ of a dissimilar second clock (practically, a clock based on another atomic species). A particular feature of these experiments is that the magnitude of the frequency change sought has to be compared to the rest frame energy $m$ of the investigated fermion (i.e., $m\sim 1$\,GeV for protons), but not to the absolute frequency $\nu$ of the clocks. It is therefore advantageous to use a very low frequency. If that is in the kHz range, where the photon energy is about $h\nu=10^{-12}$\,eV, the relative change $\delta\nu/\nu$ in the frequency is $10^{18}$ times larger than $\delta m/m$. Hence the enormous resolution that is possible in clock-comparsion experiments.

The leading systematic effect in these experiments are magnetic fields, which directly couple to the spin. While excellent magnetic shielding is thus a prerequisite for performing tests of spin-dependent effects, the most recent experiment \cite{Bearetal00} uses an additional trick to reduce their influence. A vapour cell is filled with two species of noble gas ($^{129}$Xe and $^3$He). Population inversion for both is achieved by coupling to optically pumped Rb vapour. Thus, maser oscillations of both species of noble gas at frequencies in the kHz range (see Tab. \ref{clockcomp}) are achieved. The point is that both masers depend on the same magnetic field, since they are spatially perfectly overlapped. Thus, the frequency of either can be used as a sensitive probe for stabilizing the magnetic field, while the other serves as a probe for Lorentz violation. Additionally, due to the simple structure of $^3$He nuclei, the experiment obtains limits that do not rely on the above assumptions that have been made in order to simplify the theory. The frequency change of the He maser $\delta \nu_J=-3.5\tilde b_J^n+0.012\tilde d_J^n-0.012\tilde g_{D,J}^n$ due to Lorentz violation could be calculated. Because of the relatively small numerical factors, the influence of the parameters $\tilde b_J^n, \tilde g_{D,J}^n$ can be neglected, and the experiment results in a limit $\tilde b_J^n=(6.4\pm 5.4)\times 10^{-32}$\,GeV \cite{Bearetal00,Bearetal02}. Space--based experiments of this type are also planned for the near future \cite{Bluhmetal02,Bluhmetal03}.

\begin{table}
\centering \caption[Overview of clock--comparison experiments] {\label{clockcomp} Overview of
clock--comparison experiments with sensitivity to neutron and proton mass anisotropy $\delta m$.
The quoted accuracy applies for the measurement of $\nu_1$.}
\begin{tabular}{cccccc} \\ \hline
Reference & clock 1 & $\nu_1$ & clock 2 & $\nu_2$ & $\delta m/m_{n,p}$
\\
\hline \cite{Prestageetal85} & $^9$Be & 303\,MHz & $^1$H & 23.91401\,GHz & $10^{-26}$ \\
\cite{Lamoreauxetal86} & $^{201}$Hg & 5.5\,Hz & $^{199}$Hg & 15\,Hz & $10^{-28}$ \\
\cite{Chuppetal89} & $^{21}$Ne & 1.000\,kHz & $^3$He & 9.650\,kHz &  $10^{-28}$ \\
\cite{Berglundetal95} & $^{199}$Hg & 4.0321\,Hz & $^{133}$Cs & 1.858\,kHz & $10^{-29}$ \\
\cite{Bearetal00}$^a$ & $^3$He & 1.7\,kHz & $^{129}$Xe & 4.9\,kHz & $5\times 10^{-31}$ \\
\hline
\end{tabular}\\
{\footnotesize $^a$ see also \cite{Bearetal02}}
\end{table}

A similar trick is used in 'co-magnetometer' experiments \cite{KornackRomalis05}. Since the spin-dependent Lorentz violation fields couple to matter in a similar way as magnetic fields, magnetometers (based on the precession frequency of a spin) are sensitive to these terms. The relative sensitivity to Lorentz violation and magnetic background fields may be different for different types of atoms, so simultaneous operation of two spatially overlapped magnetometers can be used to eliminate the sensitivity to the magnetic feld.

\subsubsection{Other tests}

Limits on spin dependent electron coefficients have been obtained by torsion balances with spin polarized solids (i.e., permanent magnets). One experiment yielded $\tilde b_Z^e \simeq (2.7 \pm 1.6) \times 10^{-25}m_e$ \cite{BluhmKostelecky00,Adelbergeretal99}; in a similar experiment \cite{HouNiLi03}, $((\tilde b^e_X)^2+(\tilde b_Y^e)^2)^{1/2} \leq 6.0 \times 10^{-26}m_e$ and $|\tilde b_Z^e| \leq 1.4 \times 10^{-25}m_e$ have been found.

Hydrogen spectroscopy can prospectively limit linear combinations of $\tilde b_J^e, \tilde b_J^p$,
to about $10^{-27}$\,GeV \cite{BluhmKosteleckyRussel99}. Due to the simplicity of Hydrogen, the sensitivity of these tests is accurately calculable, making this a clean test. Comparing the frequencies of hydrogen masers, \cite{Phillipsetal01} find $|\tilde b_J^p + \tilde b_J^e| \lesssim 2 \times 10^{-28}$\,GeV.

\subsection{Anomalous dispersion in interferometry}

The study of anomalous dispersion relations, although usually motivated
on the basis of a break up of Lorentz symmetry, deserves to be discussed
separately, if nothing else because of the huge interest that
has been attracted by this phenomenology in the recent
literature (see, {\it e.g.}, Refs.~\cite{Amelino-Camelia00a,Sarkar02,JacobsonLiberatiMattingly04,Amelino-Camelia04}).
Anomalous dispersion is of course a prototypical propagation effect,
and its study would immediately suggest resorting to astrophysics
contexts, where the propagation distances can be gigantic.
We here want to stress that however there is at least one class
of controlled/laboratory experiments where extremely high sensitivity
to anomalous dispersion can be achieved.
This has been stressed in Ref.~\cite{AmelinoCameliaLaemmerzahl04} on the basis
of the observation that conceivably,
if we manage to operate them at high sensitivity using two different-wavelength
beams, modern intereferometers (of the type used for gravity-wave searches)
might achieve~\cite{AmelinoCameliaLaemmerzahl04} sensitivity to $\eta \sim 1$,
for the $n=1$ case in Eq.~(\ref{displeadbis}).

\subsection{Space--time fluctuations and decoherence}

Concerning the models of Planck-scale-induced strain noise in interferometry described in Section 3.5 the quality of experimental limits is improving quickly. For example, already available data constrain $\zeta$ to be much smaller than $1$ in the random-walk scenario of Eq.~\ref{noiserw}.
This is achieved both using large ``free-mirror" laser-light
interferometers~\cite{AmelinoCamelia99,NgvanDam00,Amelino-Camelia01,Schilleretal04},
such as TAMA (soon improving with LIGO and VIRGO), and using
small-size laser-light interferometers whose mirrors are rigidly
connected~\cite{Schilleretal04}.

The fact that
we are still unable to derive a quantitative description of
spacetime fuzziness from the various quantum-gravity proposals
does not allow us to have a good intuition for the significance
of these limits. And, while theory work attempts to clarify the situation,
it appears necessary to attempt to push the limits as far as possible.

For what concerns the density-matrix Planck-scale decoherence formalism
mentioned in Section~3.5 the best limits were obtained using data from
the CPLEAR neutral-kaon experiment.
The neutral-kaon system is very sensitive to new physics because
of the delicate balance of conventional-physics scales that governs
its peculiar features. The density matrix formalism of Ref.~\cite{Ellisetal96}
involves several dimensionless parameters which one may guess to be all
obtained as roughly the ratio between the kaon mass and the
Planck scale. For some of the parameters, CPLEAR data
already exclude this possibility experimentally.

\section{Observations in astrophysics}\label{Sec:ObservationsInAstrophysics}

As mentioned, the realization that tests of some quantum-gravity effects with extremely high sensitivity could be based on observations in astrophysics has generated over these past few years a rather large interest. In turn this has resulted in the production of quite a few
general reviews of the subject, of which
Refs.~\cite{Amelino-Camelia00a,Sarkar02,JacobsonLiberatiMattingly04,Amelino-Camelia04}
are just a representative small sample. We refer the reader to these reviews (and references therein) for a detailed discussion of the relevant phenomenological proposal. Still, in order to render our own review somewhat self-contained, in this Section we comment briefly on some of the most studied methods.

\subsection{SME astrophysics}\label{SMEastrophys}

There has been a sizeable effort of constraining SME parameters using astrophysical observations of radiation. Of course the interpretation of the data on radiation emitted by distant sources as tests of LI is faced with the problem that the source as well as the space in between (that might contain gasses, other electromagnetic fields etc.) are not under control of the experimenter, and the analysis inevitably acquires a certain level of model dependence.

The distance of the source can
only be inferred from studying other radiation,
and therefore the analysis must always
compare radiation with radiation,
which in some cases renders the analysis insensitive
to spin-independent effects.
But of course the extremely long propagation distance
is very valuable for constraints on other types of effects.

The SME 4-vector $(k_{AF})^\kappa$ from the photon
sector (Tab. \ref{photoncoeff}) (that is also
expected to vanish for theoretical reasons \cite{KosteleckyLehnert01}) has been ruled out by birefringence measurements on radiation emitted by distant radio galaxies \cite{CarrollFieldJackiw90}. The idea is that $k_{AF} \neq 0$ leads to a splitting of photons in two circularly polarized modes with different phase velocities, according to a dispersion relation
\begin{equation}
|\mbox{\boldmath$k$}| = \omega\mp \frac12 \left((k_{AF})_0-|{\mbox{\boldmath$k$}}_{AF}| \cos\theta\right) \, .
\end{equation}
Here, $k=(\omega, \mbox{\boldmath$k$})$ denotes the wave vector
and ${\mbox{\boldmath$k$}}_{AF}$ the spatial components of
$k_{AF}$. The angle $\theta$ is between the
direction of ${\mbox{\boldmath$k$}}_{AF}$ and $\mbox{\boldmath$k$}$.
The $+$ and $-$ signs correspond to right- and left-handed
circularly polarized waves, respectively. As in the
Faraday effect, this causes a rotation of the
polarization of a linearly polarized plane wave by an
angle $\delta \phi=((k_{AF})_0-|{\mbox{\boldmath$k$}}_{AF}|\cos\theta)L/2$,
where $L$ denotes the distance travelled and $\theta$ is the angle between the diection of propagation and $\mbox{\boldmath$k$}$. The radio galaxies
are expected to emit radiation that is polarized either
parallel or perpendicular to their observed elongation.
A rotation of the polarization would cause a difference
between the observed polarization and elongation;
if no difference is observed, an upper limit on $k_{AF}$
can be derived. Studying 160 sources, \cite{CarrollFieldJackiw90}
one obtains $|k_{AF}|
=\sqrt{(k_{AF})^\alpha(k_{AF})_\alpha} \lesssim 10^{-42}$\,GeV.

\begin{table}
\centering
\caption{\label{photoncoeff} Effects of the SME parameters
of the photonic sector on the phase velocity of light. \# denotes
the number of degrees of freedom contained in each symbol. $\beta=v/c$
is the velocity of the experiment relative to the frame of reference
in which the parameters are defined.}
\begin{tabular}{lrlrl} \hline
Parameters & \# & makes $c$ dependent on & limit & Ref. \\ \hline
$\vec k_{AF}$ & & circular polarization & $10^{-42}$\,GeV & \\
$(\tilde\kappa_{e+})^{jk}=(\tilde\kappa_{e+})^{kj}$ & 6 & linear
polarization & $10^{-32}$ & \cite{KosteleckyMewes01,KosteleckyMewes02} \\
$(\tilde\kappa_{o-})^{jk}= - (\tilde\kappa_{o-})^{kj}$ & 3 & linear
polarization & $10^{-32}$ & \cite{KosteleckyMewes01,KosteleckyMewes02} \\
$(\tilde\kappa_{e+})^{jk}=(\tilde\kappa_{e+})^{kj}$ & 6 & direction
of propagation & $10^{-15}$ & \cite{Muelleretal03c,Wolfetal04a} \\
$(\tilde\kappa_{o+})^{jk}= - (\tilde\kappa_{o+})^{kj}$ & 3 & direction
of propagation, $\mathcal O(\beta)$ & $10^{-11}$ & \cite{Muelleretal03c,Wolfetal04a} \\
$\tilde\kappa_{tr}$ & 1 & motion
of experiment $\mathcal O(\beta^2)$ &  &  \\ \hline
\end{tabular}
\end{table}

The SME matrices $(\tilde \kappa_{e+})^{jk}$
and $(\tilde \kappa_{o-})^{jk}$ from the
photon sector (Tab. \ref{photoncoeff})
have also been restricted by polarization measurements.
A difference of the phase velocity $\Delta v_p$ of the
two polarization modes travelling over a distance $L$
would lead to a change in the relative phase
\begin{equation}
\Delta\phi\approx 2\pi \Delta v_p \frac{L}{\lambda} \, ,
\end{equation}
where $\lambda$ is the wavelength. This modifies the
polarization, e.g., light that is initially linearly
polarized would in general become elliptically polarized.
The large factor $L/\lambda$ makes the experiment very
sensitive. A change of the polarization state at the
receiver occurs as a consequence of either a change
in $L$ or $\lambda$. For a astrophysical source, $L$
is fixed, but one can measure the dependence on $\lambda$
(making the assumption that the emitted polarization is
relatively constant over a range of wavelengths).
From data on 16 sources at 0.04\,Gpc$\leq L \leq$3.53\,Gpc,
a limit of $2\times 10^{-32}$ (90\% confidence level) on
all elements of $(\tilde \kappa_{e+})^{jk}$
and $(\tilde \kappa_{o-})^{jk}$ is
obtained \cite{KosteleckyMewes01,KosteleckyMewes02}.

Similar analyses of synchrotron radiation from the Crab nebula
and other sources are reported
by \cite{JacobsonLiberatiMattingly03}. As stressed in the next subsection,
they interpret the result in terms of a modified dispersion relation of the
form (\ref{displeadbis}). Since their analysis
uses birefringence methods, it should be possible
to interpret the same result in terms of the SME:
The wavelength of the $\sim 0.3$\,MeV radiation
is $\sim 4\times 10^{-12}$\,m, and the distance
is assumed to be $L=0.5$\,GPc=$1.5\times 10^{25}$\,m
or about $10^{37}\lambda$. Since the upper limit on
polarization changes with $\lambda$ is of the order
of one radian, an analysis of these results in terms
of the SME will likely lead to a limit of the order
of $10^{-37}$ on at least one component of
either $(\tilde \kappa_{e+})^{jk}$ or $(\tilde \kappa_{o-})^{jk}$.

Another kind of astrophysical limit is derived from the following idea: If Lorentz symmetry is not exact, the limiting velocity for different particles does not need to be equal. A 'fast' particle with a higher limiting velocity may then emit Cherenkov radiation that consists of particles with lower limiting velocities. Thus, its velocity would eventually be reduced to below the threshold for Cherenkov radiation in a time that is short compared to astrophysical timescales, so particles above a certain energy would not exist in cosmic radiation. From observations of cosmic radiation with energies of up to $3\times 10^{11}$\,GeV, the maximum velocity of several particles can be constrained to deviate no more than a few parts in $10^{-20} \ldots 10^{-24}$ from $c$ \cite{GagnonMoore04}.

\subsection{Anomalous dispersion}

A special role within the phenomenology inspired by the possibility
of Lorentz-symmetry violations is played by effects of
anomalous dispersion. The fact that (as also stressed earlier in these notes)
several approaches to the quantum-gravity problem motivate
the study of Planck-scale-modified dispersion relations,
has led, over the last few years, to a very large effort
on the phenomenology side.

The first popular idea for searches of Planck-scale anomalous dispersion
was based~\cite{Amelino-Cameliaetal98}
on properties of gamma-ray bursts. These bursts travel cosmological distances and contain a rich
time structure, and as a result they can be used to set rather significant limits on the wavelength dependence of the speed of photons which would follow from a dispersion relation
of the type (\ref{displeadbis}). The present best limits obtained
using this strategy constrain
(see, {\it e.g.}, Refs.~\cite{Schaefer99,Amelino-Camelia02a,Ellisetal03a}) the parameter $\eta < 300$ for the case $n=1$ in (\ref{displeadbis}). And planned telescopes should provide an improvement of a few
orders of magnitude within a few years~\cite{Norrisetal99,deAngelis00}.

For the case in which the modified dispersion relation is assumed
to hold within a field-theoretic setup very stringent (beyond-Plankian)
limits can be obtained from an analysis of synchrotron radiation from the Crab
nebula \cite{Jacobsonetal03}.
However, the field-theoretic setup in the photon sector
imposes birefringence
and invites one to consider a particle/helicity dependence of $\eta$,
leading to many parameters. Only one of these parameters can be constrained
using the Crab-nebula synchrotron-radiation
analysis~\cite{Amelino-Camelia02,Jacobsonetal03}.

The case of the Crab-nebula synchrotron radiation is interesting
because, as stressed already in the previous subsection,
this same set of data can be valuable both for the SME and
for the phenomenology based on (\ref{displeadbis}).
Clearly the physics described by the SME matrices is
different from the energy-dependent effects~\cite{Amelino-Cameliaetal98}
associated with (\ref{displeadbis}). In principle,
simultaneous limits on both effects could be obtained by
studying the energy-dependence of the phase-shift. This illustrates that
some observation is sometimes interpreted in two
theoretical frameworks to produce results that
appear to be unrelated, while the actual observation is the same.

Still concerning the hypothesis of a description of
birefringence based on (\ref{displeadbis})
a limit of order  $|\eta| < 2 \cdot 10^{-4}$, for $n=1$,
can be inferred~\cite{GleiserKozameh01} using observations of polarized light
from distant galaxies.

Perhaps the most exciting opportunity for this phenomenology
based on (\ref{displeadbis})
is provided by the study of ultrahigh-energy cosmic rays.
Before reaching our observatories ultrahigh-energy
cosmic rays travel gigantic (cosmological) distances and
our expectations concerning the structure of the cosmic-ray
spectrum depend strongly on the
interactions of cosmic rays with photons in the cosmic microwave
background, which in particular should produce
a cut off in the spectrum around $5 \cdot 10^{19}$ eV.
A modification of the dispersion relation
could lead~\cite{Kifune99,Aloisioetal00,Amelino-CameliaPiran01} to a modification of
this cut off prediction.
The Auger cosmic-ray observatory, now starting to take data,
has the sensitivity needed to probe very small values
of $\eta$ for $n=1$, and even in the case $n=2$ Auger could still
be sensitive to $\eta$ of order 1.

\subsection{Spacetime fuzziness and decoherence}
Some proposals concerning the use of observations in astrophysics
to explore the quantum-gravity ideas of spacetime fuzziness
and decoherence have also been made.
In Ref.~\cite{LieuHillman03}
it is argued that  evidence of a good phase coherence of
light from extragalactic sources
could be used to constrain spacetime-fuzziness scenarios.
This proposal opens a
valuable phenomenological window, but
it might be necessary to seek improved analyses, taking into account
the comments made in Ref.~\cite{NgvanDamChristiansen03}
concerning the quantum-gravity aspects,
and in Ref.~\cite{Coule03}
concerning the astrophysics aspects.

Concerning decoherence certain aspects of neutrino astrophysics
may play a role~\cite{Mavromatos04} somewhat similar to the one
played by laboratory studies of the neutral-kaon system.

\section{Summary and outlook}

We have given a rather general overview of the
phenomenology work which finds motivation in the
study of the quantum-gravity problem. The field
has grown so wide that we shall not here argue
that we gave a complete overview, but we did set as
one of our primary goals the one of illustrating some
alternative strategies which are being pursued for this phenomenology.
Within this spectrum of research programmes one finds on one side a
phenomenology of the type illustrated by studies of the
modified dispersion relation (\ref{displeadbis}), with
only a minimal parametrization, a nearly exclusive focus
on the hypothesis that the onset of the new effects should
be characterized by a scale which is rather close to the
Planck scale, and the expectation that the new effects
should not be describable in terms of renormalizable
field-theory operators. And on the opposite side the
strategy involves very general parametrizations, a
phenomenology which looks for the new effects even
at scales much different from the Planck scale, but
assuming throughout that the present formalisms can
accommodate the new effects, so that one is led for
example to the SME field theory and/or to the type
of modified Maxwell and Dirac equations which we discussed.
To our knowledge there are no other reviews covering this
wide spectrum of approaches, and we thought it might be
beneficial to provide one. In spite of the differences
in strategy there are a number of common challenges for
all these approaches, besides the obvious common goal of contributing to
the search of a solution for the quantum-gravity problem.

Another key goal for our work was to stress that this candidate effects that have been discussed in the quantum-gravity literature are not merely of academic interest, as it is sometimes assumed. We believe that the example of metrology,
which we discussed in Subsection~1.2,
should act as a clear warning of the potential (even technological!) implications of these studies.

We expect that this field will prove to be among the most exciting of the next 10 or 20 years, since over this period we can clearly see several opportunities for sharp improvement of the present sensitivities toward the relevant effects. A good example is provided by the Auger cosmic-ray data which will
surely amuse us for the next couple of years, and are
going to provide a much clearer picture of the high-energy cosmic-ray spectrum. Similarly, the next generation of gamma-ray telescopes, such as GLAST, will provide a huge improvement for the relevant gamma-ray studies.

While many Lorentz-violating effects have been bounded, some to impressive levels of precision, improvements in terms of accuracy, cleanliness and comprehensiveness are expected from improved versions of the old experiments and
from experiments of new kinds. New turntable versions of the cavity tests are performed at several locations. The shorter timescale of the rotation (minutes compared to 24\,h for Earth's rotation) allows to collect much more data in a
given time and to exploit the time scale of the optimum sensitivity of the experiment. This should allow two to three orders of magnitude improvements, if the systematic effects associated with active rotation (like the bending of the resonators caused by a slight tilt of the rotation axis relative to Earth's gravity) can be controlled. As an alternative idea, cavity experiments could also search for birefringence in isotropic materials induced by Lorentz violation. Such experiments could perform the same tests as conventional Michelson-Morley type experiments, but would be immune against a
broad range of systematic effects,
including length changes of the cavity \cite{Mueller04}.

A very interesting challenge is to put some of these experiments into space.
This would have several advantages. The absence of seismic vibrations is
important for experiments involving macroscopic matter. Experiments in
space also offer the possibility to choose the time scale of rotation
and orbit of the satellite which provide the necessary modulation of
the Lorentz violation in the frame of the experiment. On Earth, both
time scales (24\,h and 365\,d) are relatively long. In space, these
would be reduced to minutes and hours without introducing the systematic
effects of turntable experiments on Earth.
Several proposals \cite{Bluhmetal03,Laemmerzahletal04a,Buchmanetal00}
have been made for experiments on the International Space Station ISS and
dedicated satellites.

New types of terrestrial experiments are
suggested by new theoretical work. For example,
the relatively low accuracy of the bounds on the
photon parameters $\tilde \kappa_{0+}$ can be
improved by exploring the static limit of
Lorentz-violating Maxwell
equations \cite{LaemmerzahlMaciasMueller05,BaileyKostelecky04}.
Such experiments can also bound the Lorentz-violating non-SME terms of the most general Maxwell equations \cite{LaemmerzahlMaciasMueller05}.

Concerning the UFF, we expect improvements by one order of magnitude within the next few years for laboratory experiments and up to five orders with dedicated satellite missions. These experiments, when carried through successfully, will definitely for the first time decide on the viability of certain quantum gravity predictions.

\section*{Acknowledgments}

We like to thank H. Dittus, F.W. Hehl, A. Kosteleck\'{y}, A. Peters, and S. Schiller for fruitful discussions and ongoing cooperation. Financial support from the German Space Agency DLR, the German Academic Exchange Service DAAD, the Alexander von Humboldt-Stiftung, and CONACYT Grant 42191-F is acknowledged.

\appendix
\section{Standard Model Extension definitions and conventions}

Rich and extensive literature exists on the SME, so a full description is neither necessary nor possible here. However, we introduce briefly the definition and standard notation for Lorentz violating quantities in the SME.

\paragraph{Lagrangian}

The SME starts from a Lagrangian formulation of the Standard Model, adding 
all possible observer
Lorentz scalars that can be formed from the known particles and Lorentz 
tensors. Taken from the full SME that includes all known particles, the 
Lagrangian involving the Dirac field $\psi$ of one fermion and the 
electromagnetic field $F^{\mu\nu}$ can be written as
\begin{eqnarray}
{\mathcal L} & = & \frac i2 \bar \psi \Gamma_\nu D^\nu \psi -\frac 12 \bar 
\psi M
\psi + \mbox{h.c} \nonumber \\ & &  -\frac14 F^{\mu\nu}F_{\mu\nu}
-\frac14(k_F)_{\kappa\lambda\mu\nu}F^{\kappa\lambda}F^{\mu\nu}+\frac12(k_{AF})^\kappa\epsilon_{\kappa
\lambda\mu\nu}A^\lambda F^{\mu\nu} \, ,
\end{eqnarray}
where h.c. denotes the Hermitian conjugate of the previous terms, and 
$A^\lambda$ is the vector
potential. The symbols $\Gamma_\nu$ and $M$ are given by
\begin{eqnarray}
\Gamma_\nu & = & \gamma_\nu + c_{\mu \nu} \gamma^\mu + d_{\mu \nu} \gamma_5 
\gamma^\mu+ e_\nu
+if_\nu \gamma_5 + \frac 12 g_{\lambda \mu \nu} \sigma^{\lambda \mu} \, , 
\nonumber \\ M & = & m +
a_\mu \gamma^\mu + b_\mu \gamma_5 \gamma^\mu + \frac 12 H_{\mu \nu} 
\sigma^{\mu \nu} \, .
\end{eqnarray}
The SME introduces such parameters into the Lagrangian of each type of 
fermion, as denoted by superscripts added to the field $\psi$ as well as to 
the symbols $a_\mu$, $b_\mu$, $c_{\mu \nu}$,
$d_{\mu\nu}$, $e_\mu$, $f_\mu$, $g_{\lambda \mu \nu}$, and $H_{\mu \nu}$. 
The $\gamma_\nu, \gamma_5$ and $\sigma^{\mu \nu}$ are the conventional Dirac 
matrices, and $D^\nu$ is
the usual covariant derivative. The tensors entering $M$ have the dimension 
mass, the others are dimensionless. $H_{\mu \nu}$ is antisymmetric; 
$g_{\lambda \mu \nu}$ is antisymmetric in its first two indices. $c_{\mu 
\nu}$ and $d_{\mu \nu}$ are traceless. Gauge invariance and 
renormalizability excludes $e_\nu, f_\nu$, and $g_{\lambda \mu \nu}$, so 
these may be assumed to be either zero or suppressed relative to the other 
terms. Analogous terms are obtained in the framework of the generalized 
Dirac equation described in Sec.\ref{Sec:GeneralizedDirac}.

Lorentz violation for the photons is encoded in the tensors 
$(k_{AF})^\kappa$ and
$(k_F)_{\kappa\lambda\mu\nu}$. They correspond to the terms 
$\chi^{\kappa\lambda\mu\nu}$ and $\chi^{\kappa\lambda\mu}$ of the 
non-Lagrangian framework described in Sec.\ref{Sec:GeneralizedMaxwell}.

\paragraph{Coordinate and field definitions}

Some of the Lorentz violating parameters contained in one sector of the SME 
can be absorbed into the other sectors by coordinate and field 
redefinitions. For example, in a hypothetical world containing only photons 
and electrons, nine components of $(k_F)_{\kappa\lambda\mu\nu}$ could be 
moved into the nine symmetric components of $c_{\mu\nu}$. By definition, 
either the photon or the electron sector could be taken as conventional with 
respect to these parameters, with the Lorentz violation in the other sector. 
The presence of other particles, of course, changes this picture. We can 
still assume that one of the sectors is conventional, but then in general 
the other sectors are Lorentz-violating. Loosely speaking, in experiments 
where one compares the sectors against each other, only differential effects 
are meaningful.


\end{document}